\documentclass[11pt,a4paper]{article}

    \usepackage{jheppub}
	\usepackage[utf8]{inputenc}
	\usepackage{setspace}
	\usepackage{pdfpages}
	\usepackage{lmodern}
	\usepackage{xcolor,color,soul,colortbl}
	\usepackage{graphicx,graphics}
	\usepackage{indentfirst}
	\usepackage{amsmath,amssymb}
	\usepackage{multirow,bigstrut}
	\usepackage{epstopdf}
	\usepackage{psfrag}
	\usepackage{pstool}
	\usepackage{slashed}
	\usepackage{bbold}
	\usepackage[makeroom]{cancel}
	\usepackage{xspace}
	\usepackage{rotating}
	\usepackage[normalem]{ulem}
	\usepackage{mathtools}
	\usepackage{enumitem}
	\usepackage{subcaption}

	\graphicspath{{./Images/}}

\newcommand{\PH}{\ensuremath{h}\xspace}
\newcommand{\PZ}{\ensuremath{Z}\xspace}
\newcommand{\ttbar}{\ensuremath{t\overline{t}}\xspace}
\newcommand{\ttH}{\ensuremath{\ttbar\PH}\xspace}
\newcommand{\ttW}{\ensuremath{\ttbar W^\pm}\xspace}
\newcommand{\ttZ}{\ensuremath{\ttbar Z}\xspace}
\newcommand{\ttHH}{\ensuremath{\ttbar\PH\PH}\xspace}
\newcommand{\ttZH}{\ensuremath{\ttbar\PZ\PH}\xspace}
\newcommand{\tttt}{\ensuremath{\ttbar\ttbar}\xspace}
\newcommand{\ttWW}{\ensuremath{\ttbar W^+W^-}\xspace}
\newcommand{\Wleppm}{\ensuremath{W_{l^\pm}}}
\newcommand{\Wlepmp}{\ensuremath{W_{l^\mp}}}
\newcommand{\Whad}{\ensuremath{W_\text{had}}}
\newcommand{\Zlep}{\ensuremath{Z_l}}
\newcommand{\Zhad}{\ensuremath{Z_\text{had}}}
\newcommand{\Zb}{\ensuremath{Z_{b \bar{b}}}}
\newcommand{\WWttH}{\ensuremath{W^+W^-\ttbar\PH}\xspace}
\newcommand{\WWttZ}{\ensuremath{W^+W^-\ttbar\PZ}\xspace}
\newcommand{\X}{\ensuremath{X_{5/3}\xspace}}
\newcommand{\five}{MCHM$_5$\xspace}
\newcommand{\Tone}{\ensuremath{T^{(1)}\xspace}}
\newcommand{\Tonebar}{\ensuremath{\bar{T}^{(1)}\xspace}}
\newcommand{\Ttwo}{\ensuremath{T^{(2)}\xspace}}
\newcommand{\Tthree}{\ensuremath{T^{(3)}\xspace}}
\newcommand{\ttZZ}{\ensuremath{\ttbar\PZ\PZ}\xspace}
\def\be{\begin{equation}}
\def\ee{\end{equation}}
\def\bea{\begin{eqnarray}}
\def\eea{\end{eqnarray}}
\def\beann{\begin{eqnarray*}}
\def\eeann{\end{eqnarray*}}

\newcommand\T{\rule{-3.7pt}{3ex}}       
\newcommand\B{\rule[-1.5ex]{-3.7pt}{0pt}} 
\newcommand\TB{\rule[-1.5ex]{-3.7pt}{4.5ex}}

\makeindex
\begin{document}

\title{On the Importance of Three-Body Decays of Vector-Like Quarks}

\author{Carlos Bautista$^{1}$,}

\author{Leonardo de Lima$^2$,}

\author{Ricardo D'Elia Matheus$^1$,}

\author{Aurore Savoy-Navarro$^3$}

\affiliation[1]{Instituto de F\'isica Te\'orica, UNESP, S\~ao Paulo, Brazil}


\affiliation[2]{Universidade Tecnol\'ogica Federal do Parana, Toledo, Brazil}

\affiliation[3]{IRFU-CEA, Universit\'e Paris-Saclay and CNRS-IN2P3, France}

\abstract{It is a common feature of vector-like extensions of the electroweak sector to have near degenerate states, such as electroweak doublets. In simplified models, it is usually assumed that these have decay widths saturated by two-body channels. As a consequence, experimental searches can be done focusing on only one of the states of the doublet. 
Taking as an example case the light exotic electroweak doublet present in the Minimal Composite Higgs Model, we show that including three-body decays in the pair production process makes this separation unfeasible, since both states of the doublet will be present and contribute significantly to the signal. In addition, by recasting present searches in multileptonic channels, with a simplified cut-and-count analysis, a relevant increase in discovery reach or exclusion potential is obtained; this indeed motivates
a more detailed analysis. This study shows how
an inclusive search strategy, taking into account both the near degeneracy and the presence of three-body decays, will have greater discovery power and be more natural from a model building perspective.}

\newpage
\maketitle
\section{Introduction}
Vectorlike quarks (VLQs) are a common feature of many models of physics beyond the Standard Model (SM), aiming to naturally obtain a hierarchy between the electroweak scale and new physics at the TeV scale. Models such as composite Higgs models~\cite{GEORGI1984216, Kaplan:1983fs, KAPLAN1984187, Dugan:1984hq}, warped extra-dimensional models~\cite{Randall_1999_1, Randall_1999_2} and Little-Higgs models~\cite{PhysRevD.10.539, PhysRevD.12.508, Arkani-Hamed:2001nha, Arkani_Hamed_2002_littlest, Arkani_Hamed_2002_moose} implement a composite strongly coupled sector as the high energy completion of the SM, with the Higgs doublet being constructed from pNGBs from a dynamical symmetry breaking  happening at some UV scale (beyond a few TeV). In this kind of dynamical models, fermion masses are generated by higher dimensional operators that mix the SM fermionic sector with the strong sector, in a scheme called partial compositeness~\cite{KAPLAN1991259}. The resulting mass spectrum is composed of the lighter SM chiral fermions and heavier vectorlike partners. 
The first two families of quarks and leptons are expected to have a small mixing with the strong sector, both from the point of view of theory
and experiment~\cite{Panico:2015jxa}, as their masses lie far below the EW scale,
so viable models have the partners of these fermions well into the UV, if at all present. The same is not true for the third generation, as naturalness favors light top partners \cite{Matsedonskyi_2013} and current constraints allow for the existence of these states around 1.5 TeV (the exact constraint depending on the model). 

Here we will focus on VLQs arising in the Minimal Composite Higgs Model (MCHM)~\cite{Agashe:2004rs}, that obtains the EW doublet as the pNGB of the $SO(5)/SO(4)$ breaking pattern, and preserves custodial symmetry. The model has been comprehensively reviewed in~\cite{Panico:2015jxa} and we will not cover it in detail. For our purposes it suffices to know that the strong sector fermions fit into complete representations of $SO(4)$ that we can group together to form representations of $SO(5)$. From the point of view of phenomenology, that means that top partners usually do not come alone, with some considerable tuning needed to push most of the new vectorlike states away from the lightest one. Even in the simplest embedding, that consists of a $SO(4)$ fourplet and a $SO(4)$ singlet, there are five vectorlike states: two top partners, a bottom partner, and two exotic states with hypercharge $7/6$ and electric charges of $2/3$ and $5/3$. As we review below, for most points in the parameter space at least two of these states (an electroweak doublet) will be degenerate or near-degenerate in mass, and in many cases more than two will be close together. An important result of~\cite{Bautista:2020mxw} is that for a big part of the parameter space the top partner has sizeable 3-body decays.

The direct experimental searches for vectorlike top partners and exotic VLQs, on the other side, have focused mostly on model independent searches based around two main assumptions~\cite{Gripaios_2014,Backovi__2016, CMS:2019eqb, Sirunyan:2018omb, Aaboud:2018pii, PhysRevD.98.112010, x53search2014, x53search2017, x53search2019}:

\begin{enumerate}[label=(\alph*)]
    \item There is only one VLQ contributing to the signal chosen. Other BSM resonances are much heavier, absent or decay into different final states. Separate searches are carried out for the two most popular VLQs: the top-partner $T$ and the exotically charged \X .
    \item The decay width is saturated by a few 2-body decay modes of the VLQs. Specifically, the $\X$ is assumed to decay only through $\X \rightarrow t W^{+}$ and the T has three decay modes: $T \rightarrow b W^{+}$, $T \rightarrow t Z$ and $T \rightarrow t h$, with searches making different assumptions on the branching ratios of these three channels, but always considering that they add up to one.
\end{enumerate}

These two assumptions have an important interplay, as limiting the decays to 2-body channels is what allows the $T$ and the $\X$ to be searched for separately, even if they are close in mass. We will show that as soon as one considers 3-body decays both resonances will contribute to the same final states.

The aim of this work is thus to evaluate the impact of considering the typical situation of complete models for VLQs, which in general violate assumptions (a) and (b) above. In section~\ref{recast} we recast existing searches by relaxing assumption (b) and allowing for 3-body decay channels, obtaining an estimate on how much the exclusion limits for the top partners $T$ and $\X$ are expected to independently change. In section~\ref{searchstrat} we relax also assumption (a) which together with 3-body decays means that many VLQ resonances can contribute to the same signal. In this case the searches for $T$, $\X$ and other VLQs are not independent anymore and, using a typical point in the parameter space of the \five, we propose an inclusive search strategy for new physics signals associated with the \five. We summarize our results in section~\ref{conclusion}.

\section{Effects of a three body decay channel in VLQ searches}
\label{recast}
The usual searches of vectorlike quarks assume that they have only 2-body decay channels~\cite{CMS:2019eqb, Sirunyan:2018omb, Aaboud:2018pii, PhysRevD.98.112010, x53search2014, x53search2017, x53search2019}. Here, we make a rough estimation of the effect of the inclusion of an additional three body channel to the decays of the lightest top partner $\Tone$ (so named to differentiate it from other top partners present in the MCHM) and the exotically charged $\X$ in regards to the mass exclusion for those states. Our strategy will be to follow as closely as possible the experimental analyses used to search for the pair production of both resonances, specifically those in references~\cite{tsearch} and~\cite{xsearch}, and apply the following simple steps:

\begin{itemize}
    \item simulate the pair production and decay, for a set of combinations of branching ratios, including those used in the experiments and adding new ones, with 3-body decays;
    \item apply cuts and detector simulation that are as close as possible to that used by experiments, in order to obtain the total number of signal events for each choice of branching ratios;
    \item obtain ratios between number of events in different scenarios and use those ratios to recast the existing limits to the masses of $\Tone$ and $\X$.
\end{itemize}

The strategies applied to
the $T$ and $\X$ experimental searches are similar, and briefly summarized here. The detailed description  
can be found in the references~\cite{tsearch,xsearch}.
First, a set of regularly spaced values for the mass of the VLQ is chosen. For each mass value, 
a sample of pair-produced VLQ is generated at Leading Order (LO) with MADGRAPH5 aMC@NLO. For each considered VLQ 2-body decay, the generator is interfaced with PYTHIA 8 for parton showering and fragmentation and including the final states decay into the foreseen signatures. The simulated search (signal) samples are then processed through the full GEANT-based detector simulation. This is repeated for every 2-body channel under consideration. Furthermore, for each search channels, defined by their final signal signatures the related SM backgrounds are likewise simulated after processed with different corresponding generators. The samples are then normalized using a next-to-next to leading order (NNLO) calculation of the cross section and the different decay channels are weighted to reflect the choices of branching ratios. 

For each search channel a cut-based pre-selection followed by a statistical based method and/or neural network (NN) analysis are applied to both the simulated signal and backgrounds data. This processing chain performs the event selection, categorization and reconstruction, enhancing signal to background ratio. Once the analysis with simulated data is validated, the same analysis strategy is applied to the real data. A statistical method comparison between real data and simulated ones is performed. This allows to determine the upper limit to the cross section of pair production of the VLQ at 95\% confidence level, for each value of the VLQ mass. Since the pairs are produced through standard QCD interactions, the only new physics parameter controlling the production cross section is the VLQ mass, so the upper limit on the cross section is directly converted to a lower limit on the resonance mass.

In the \five the VLQs are obtained from the mixing of the elementary fields $q_L = (t_L, b_L)$ and $t_R$ (having the same transformations under the SM gauge group as SM quarks) with the composite resonances embedded in a fiveplet of $SO(5)$ that decomposes under $SO(4)$ as a fourplet, $\Psi_4$, and a singlet, $\Psi_1$:
\bea
\Psi_4 &\sim& (X_{5/3}, X_{2/3}, T, B)~,
\nonumber \\ [0.5em]
\Psi_1 &\sim& \tilde{T}~,
\label{comp5content}
\eea
where the $T$, $B$ and $\tilde{T}$ transform as $t_L$, $b_L$ and $t_R$ respectively and the $X_Q$ are exotic states with hypercharge $Y = 7/6$ and electric charge $Q$. Up to electroweak symmetry breaking effects, the masses of the resonances are:
\begin{equation}
\label{eq:mass}
M_{X_{Q}} = |M_4|,~M_{T, B} = \sqrt{M_4^2+y_L^2 f^2}; ~ M_{\tilde{T}} = \sqrt{M_1^2+y_R^2 f^2}, 
\end{equation}
where $M_{1,4}$ are the vectorlike masses of $\Psi_{1,4}$ and $y_{L,R}\, f$ controls the strength of the mixing of the resonances with $t_{L,R}$. See \cite{Bautista:2020mxw} for further details. From these expressions it is clear that the doublets are near degenerate, as we stated before.
The diagonalization of the charge 2/3 mass matrix, involving the states $t$, $X_{2/3}$, $T$ and $\tilde{T}$, will produce the experimentally observed chiral top quark and three vectorlike top partners, which we denote by $\Tone$, $\Ttwo$ and $\Tthree$ in order of increasing mass. From the approximate expressions of Eq. (\ref{eq:mass}), one sees that $\Tone$ is typically composed of mostly $X_{2/3} \subset \Psi_4$ if $|M_4|< \sqrt{M_1^2+y_R^2 f^2}$ or $\Tilde{T} \subset \Psi_1$ otherwise. The three body decays of $\Tone$ are also highly dependent on its composition:
\be\label{lin_comb_T1_L}
T^{(1)}_L=U_{L,1} t_L + U_{L,2}T_L + U_{L,3}X_{2/3 L} + U_{L,4}\tilde{T}_L
\ee
\be\label{lin_comb_T1_R}
T^{(1)}_R=U_{R,1} t_R + U_{R,2}T_R + U_{R,3}X_{2/3 R} + U_{R,4}\tilde{T}_R
\ee
with $L$ and $R$ indicating the chiralities of each state,and $U_{L,R}$ the corresponding unitary rotations to the mass basis. We define:
\begin{equation}
    \sin^2 \theta = \frac{\eta_L^{F}+\eta_R^{F}}{2} \;\;\;\mbox{and}\;\;\; \cos^2 \theta=\frac{\eta_L^{S}+\eta_R^{S}}{2},
    \label{fsmix}
\end{equation}
where $\eta^F_{L(R)}$ and $ \eta^S_{L(R)}$ are respectively the fourplet and singlet contributions for each chirality:
\begin{equation}
    \eta^F_{L(R)}=U_{L(R), 2}^2+U_{L(R), 3}^2
\end{equation}
 \begin{equation}
     \eta^S_{L(R)}=U_{L(R),1}^2+U_{L(R),4}^2
 \end{equation}

\begin{figure}
\includegraphics[width=0.49\textwidth]{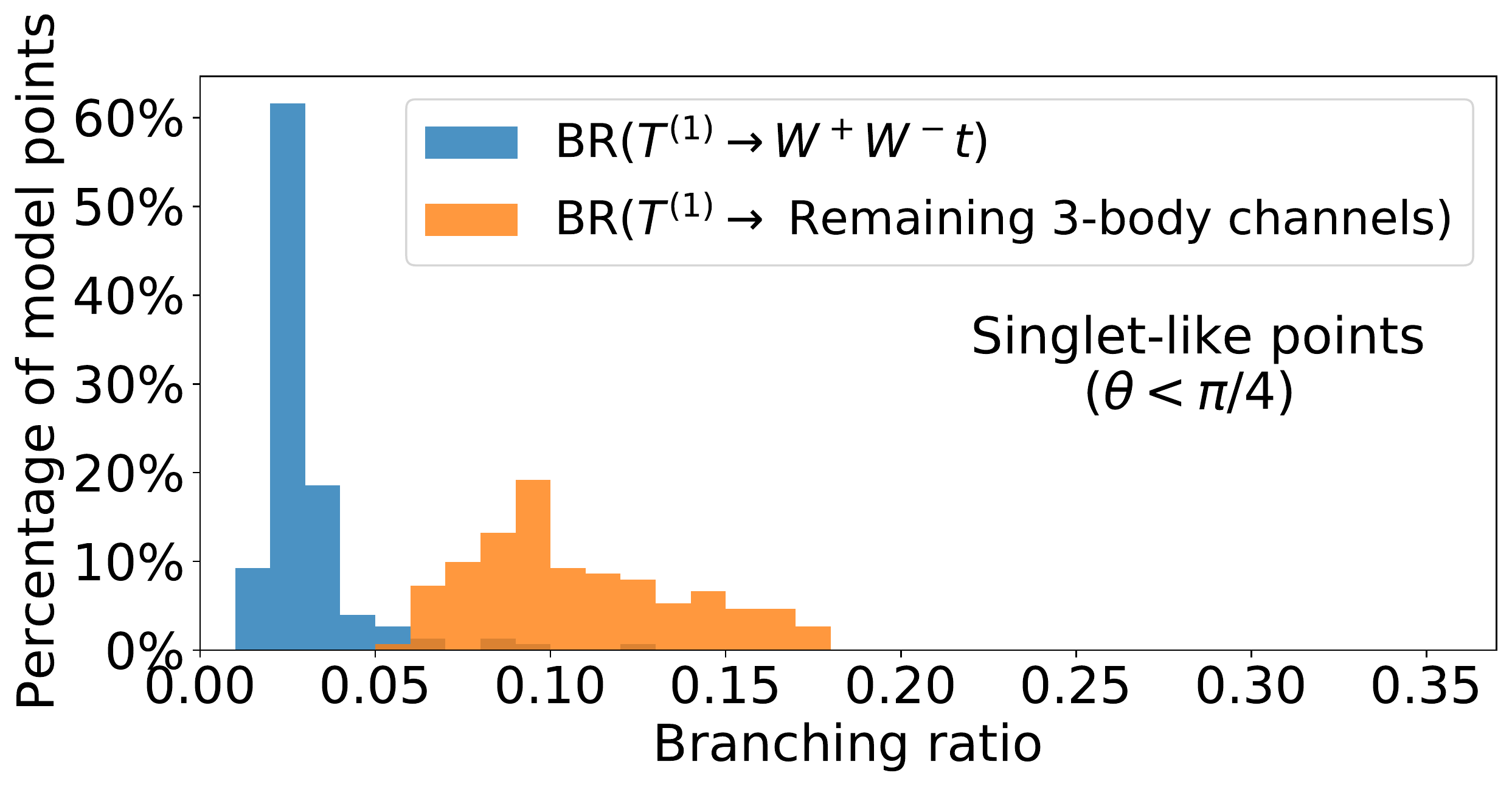}
\hspace{0.2cm}
\includegraphics[width=0.48\textwidth]{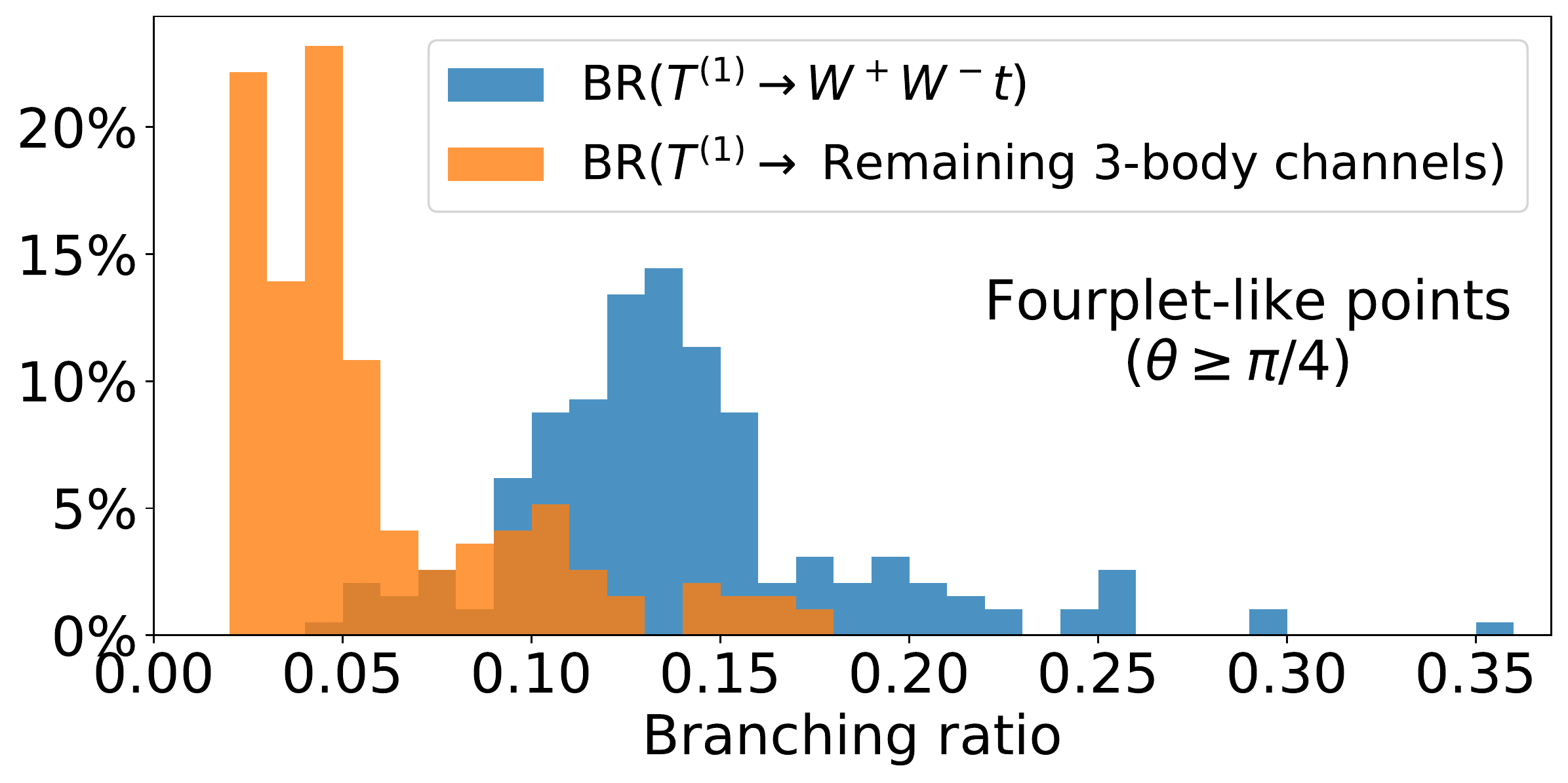}
\caption{Distributions of the branching ratio of the $T^{(1)}\rightarrow W^+W^-t$ channel (in blue) and the sum of the remaining three body decay channels (in orange). The remaining three body channels are $\bar{t}tt$, $\bar{b}hW^+$, $\bar{b}ZW^+$, $b\bar{b}t$, $hht$, $hZt$ and $ZZt$.}
\label{fig:fvss}
\end{figure}

The angle $\theta$ in~(\ref{fsmix}) characterizes the nature of $\Tone$, with $\theta = \pi/2$ being a pure fourplet and $\theta = 0$ a pure singlet. We will divide our parameter space in two regions by saying that $\Tone$ is \emph{fourplet-like} if $\theta \geq \pi/4$ and is \emph{singlet-like} if $\theta < \pi/4$. In order to study the behaviour of three body decays in these two regions we scan over $0.8~ \mathrm{TeV} \leq f \leq 2 ~\mathrm{TeV} , 1~ \mathrm{TeV} \leq|M_{1,4}| \leq 3 ~\mathrm{TeV},  0.5\leq y_L \leq 3$, with $y_R$ fixed by the top mass, and select from these the points that pass experimental constraints, as detailed in \cite{Bautista:2020mxw}\footnote{Using the free phases of the fields, we may take all parameters except for one to be positive, which we take to be $M_1$ \cite{Bautista:2020mxw}.}. The results are shown in figure~\ref{fig:fvss}.

Figure~\ref{fig:fvss} shows that fourplet-like $\Tone$ will on average have a bigger branching ratio on 3-body decays with a prevailing decay in  $T^{(1)}\rightarrow W^+W^-t$, while the singlet-like case is the opposite, with no clear prevalence of any channel and generally smaller 3-body decay branching ratios. This motivates us to focus on the fourplet-like case in what follows. 

\begin{figure}
\centering
\includegraphics[width=0.5\textwidth]{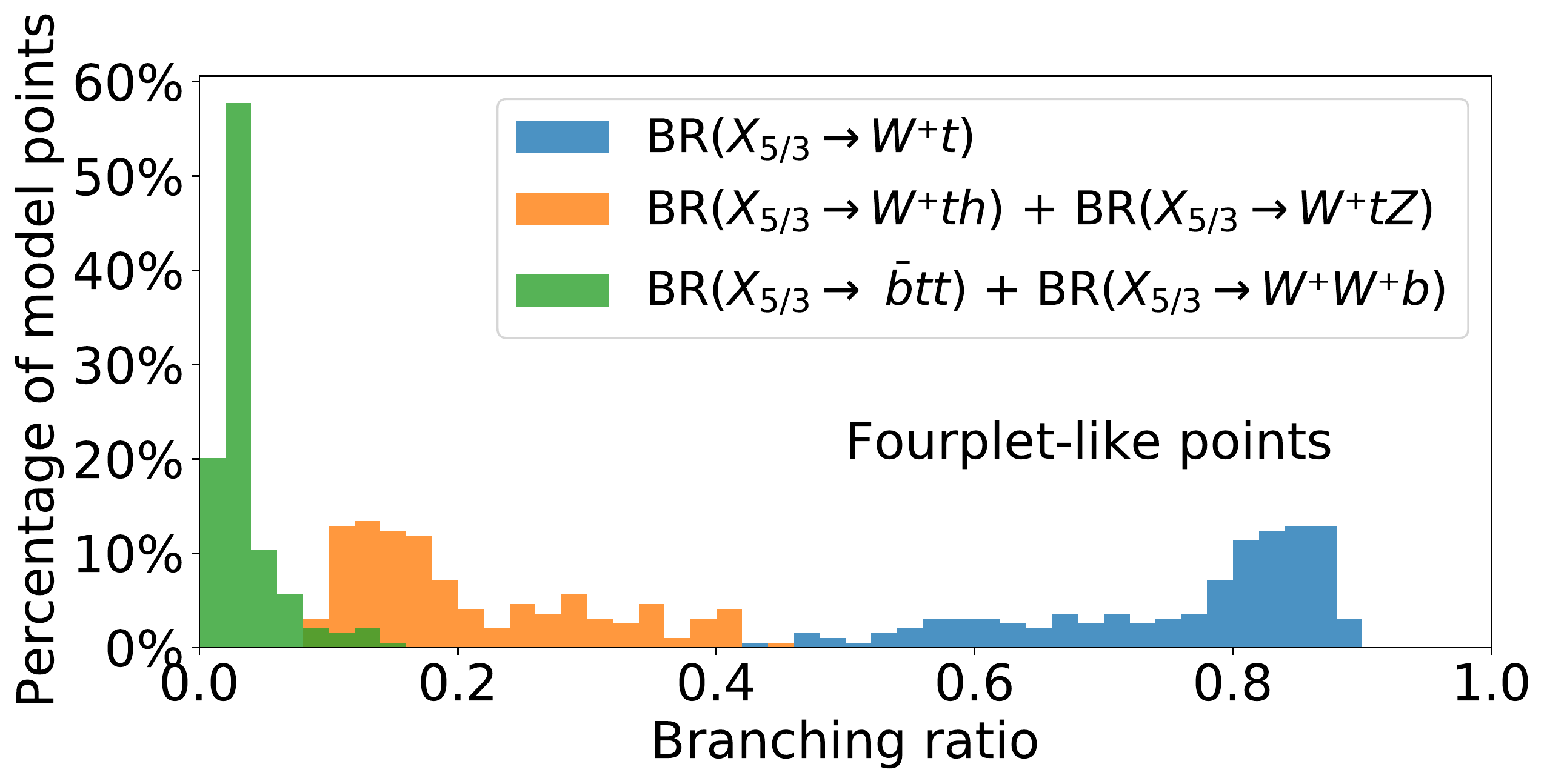}
\caption{Distribution of the branching ratios of the 5/3 charged resonance ($X_{5/3}$) decays for fourplet-like points.}
\label{fig:br_X53_MCHM5_fourplet_points}
\end{figure}

The fourplet-like scenario is also more interesting for the $\X$, as it will be one of the lightest states and in fact near degenerate with $\Tone$, with splitting caused by electroweak effects and smaller than $m_W$. In this case, the only allowed 2-body decay is $\X \rightarrow W^+ t$. The possible 3-body decays are listed in figure~\ref{fig:br_X53_MCHM5_fourplet_points} which shows the distribution of $\X$ branching ratios for the same model points used before. One can see that the 3-body decays can be sizeable and dominated by two channels: $X_{5/3}\rightarrow W^+th$ and $X_{5/3}\rightarrow W^+tZ$.

\subsection{Effect on the $\Tone$ search}
\label{T1recast}

\begin{figure}[h]
\centering
\includegraphics[width=0.3\textwidth]{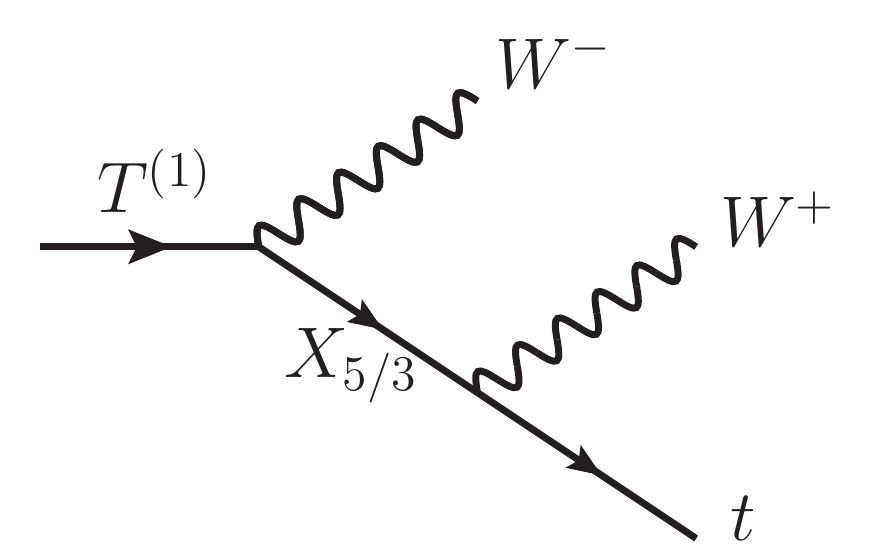}
\includegraphics[width=0.3\textwidth]{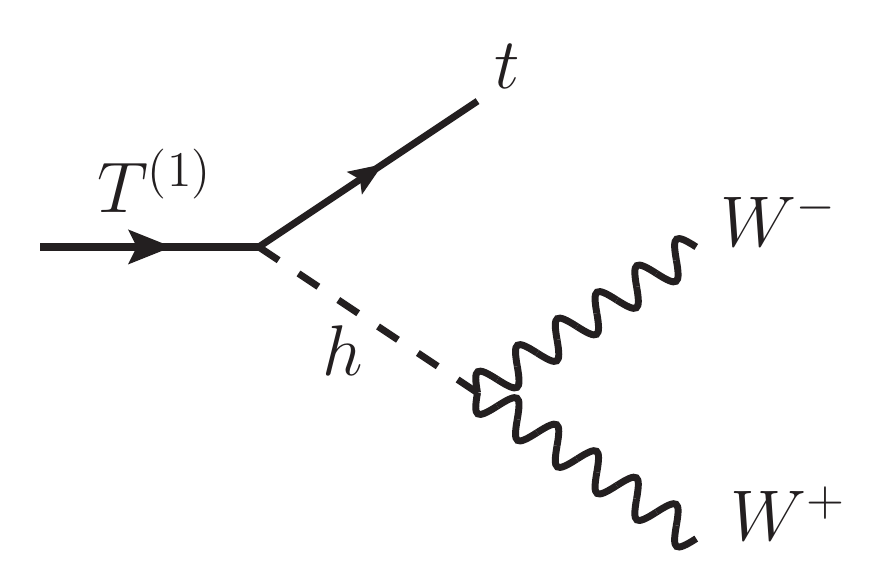}
\includegraphics[width=0.3\textwidth]{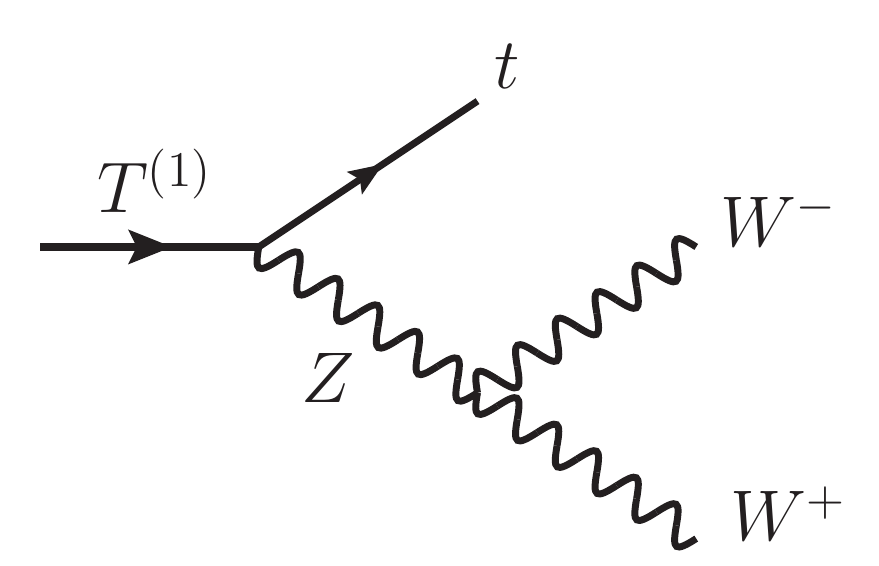}
\caption{Feynman diagrams of the main contributions to $\Tone \rightarrow W^+W^-t$ in fourplet-like scenarios.}
\label{fig:TpTOwwt}
\end{figure}

In this section we will focus on the $W^+W^-t$ decays of the fourplet-like $\Tone$ and the Feynman diagrams that mainly contribute\footnote{There are also decay channels through intermediate b and B, but those are negligible in the fourplet-like case} to the decay are shown in figure~\ref{fig:TpTOwwt}. The three channels in figure~\ref{fig:TpTOwwt} have contributions of the same magnitude and interfere positively  to increase the total three-body decay width. Regarding the two-body decays, the fourplet-like $\Tone$ has Br$\left[\Tone \rightarrow b W^{+}\right] \sim 0$, and $\mbox{Br}\left[\Tone \rightarrow t Z \right] \sim \mbox{Br}\left[\Tone \rightarrow t h \right]$.

In order to estimate the effect of the three-body decay to the existing $\Tone$ search we simulate $p\bar{p}\rightarrow \Tone\Tonebar$ for the same set of masses used in~\cite{tsearch}, choosing $m_{\Tone}$ in the range $\left[0.9 \mbox{ TeV}, 1.8 \mbox{ TeV}\right]$ in steps of 100 GeV. The pair production is followed by inclusive decays into all relevant two and three body channels, namely: $th$, $tZ$, $Wb$ and $WWt$. The events are then showered and hadronized in Pythia and finally passed to Delphes for a fast detector analysis%
\footnote{
  we used the default CMS card with few modifications in their reconstruction algorithms to follow more closely what was done in~\cite{tsearch}. The jet reconstruction was made using the anti-KT(AKT) algorithm with a radius of 0.4 and only jets with $p_T>30$ GeV and $|\eta|<2.4$ were selected and there is a requirement of isolation for leptons varying with momenta. For details see~\cite{carlosPHD}.
}. 
All simulations are done at LO, but the final cross section in each channel is rescaled to reflect a particular combination of branching ratios:
\begin{equation}
    \sigma_{p\bar{p}\rightarrow \Tone\Tonebar\rightarrow D_1\bar{D}_2}=\sigma_{p\bar{p}\rightarrow\Tone\Tonebar} \times F\left[\text{BR}(D_1), \text{BR}(D_2)\right]
\end{equation}
where $D_1$ and $D_2$ label the possible decay channels and:
  \begin{equation}\label{br_factor}
    F\left[\text{BR}(D_1), \text{BR}(D_2)\right] = 
    \begin{cases*}
      \left[\text{BR}(D)\right]^2 &, if $D_1=D_2=D$ \\
      2\times\text{BR}(D_1)\times\text{BR}(D_2)  &, if $D_1 \neq D_2$
    \end{cases*}
  \end{equation}
One can then analyze different scenarios, we will focus on the three possibilities listed in table~\ref{T_scenarios}, where the first two rows are the ones used in~\cite{tsearch} and the third is the typical fourplet-like behaviour in the \five (for $\Tone$ masses around $1.5$ TeV, $\text{BR}(W^+W^-t)$ can be larger for higher masses~\cite{Bautista:2020mxw}).

\begin{table}
  \centering
  \begin{tabular}{c|cccc}
     Scenario &  $\text{BR}(W^+b)$ & $\text{BR}(th)$ & $\text{BR}(tZ)$ &  $\text{BR}(W^+W^-t)$ \\
    \hline
    \text{``Simplified singlet''}& $0.5$ & $0.25$ & $0.25$ & $0$\\
    \text{``Simplified doublet''} & $0$ & $0.5$ & $0.5$ & $0$ \\
    \text{``Fourplet-like''} & $0$ & $0.45$ & $0.45$ & $0.1$ \\
    \end{tabular}
    \caption{Scenarios considered in the $\Tone$ analysis and their corresponding branching ratio configurations.}
    \label{T_scenarios}
\end{table}

\begin{figure}
\centering
\includegraphics[width=0.8\textwidth]{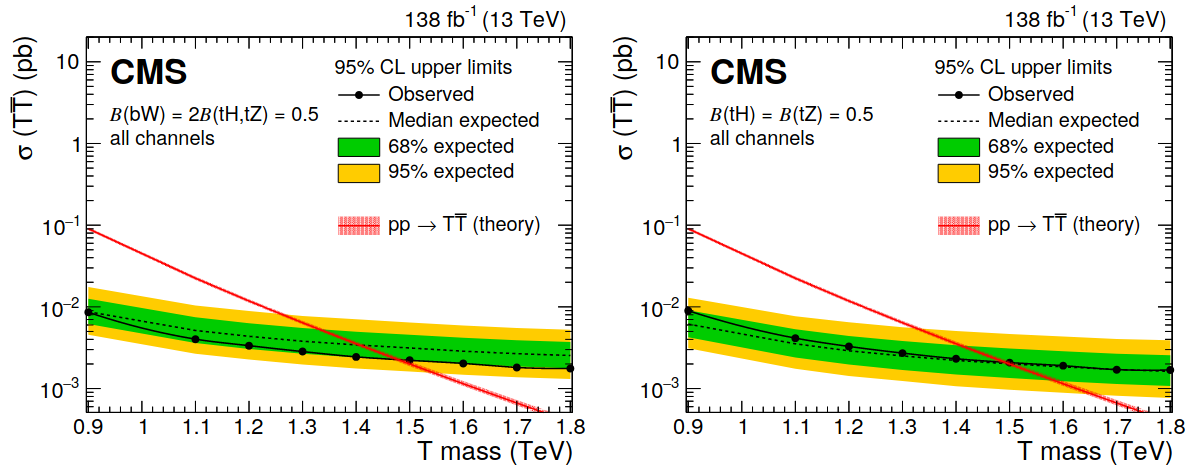}
\caption{Expected and observed limits of the signal cross section upper limit at 95$\%$ CL for the simplified singlet (left) and simplified doublet (right) scenarios obtained by combining the analyses done in~\cite{tsearch} from single lepton, same-sign dilepton and multilepton channels. The band around the theoretical prediction shows the theoretical uncertainty. Figure extracted from~\cite{tsearch}.}
\label{fig:cms_results_T}
\end{figure}

In~\cite{tsearch} three signal channels are considered: single-lepton, same-sign dilepton (2SSL) and multilepton, and the combined constraint is shown in figure~\ref{fig:cms_results_T}.  Here we focus on the 2SSL channel, where a more straightforward cut-and-count analysis was performed by CMS. It is important to understand that, since all 2-body and 3-body decays of a pair of $\Tone$ can contribute to those channels, the main effect of changing the branching ratios comes from the fact that some decay channels may be more ``resistant'' to the cuts in the analysis, specially the selection on the number and charge of leptons. The number of surviving events will be given by:
\begin{equation}
    N^{\text{CUT}} =
    \sum_{D_1, D_2}  \sigma_{p\bar{p}\rightarrow \Tone\Tonebar\rightarrow D_1\bar{D}_2}\times\xi_{D_1,D_2}\times\mathcal{L},\label{events_after_cuts}
\end{equation}
where $\mathcal{L}$ is the luminosity and $\xi_{D_1,D_2}$ is a product of detector and cuts efficiencies for each channel. 

In the 2SSL channel, exactly two isolated leptons with the same sign of electric charge are demanded. With the following cuts\footnote{
    All cuts follow ref~\cite{tsearch} and are justified there. $H^{\text{lep}}_T$ is cut at different values for different datasets used in their analysis, here we use the value for the 2017-2018 data which contains most of the analyzed luminosity.
}:
\begin{itemize}
    \item leading(subleading) lepton: $p_T > 40(30)$ GeV;
    \item all leptons: $|\eta|<2.4$;
    \item invariant mass of the 2SSL pair $m_{ll} > 20$ GeV and outside the Z window: $[76.1 \text{ GeV},\\ 106.1 \text{ GeV}]$;
    \item number of jets $N_j \geq 4$ (AKT with $R = 0.4$, $p_T > 30$ GeV and $|\eta| < 2.4$);
    \item $H^{\text{lep}}_T> 400 \text{ GeV}$.
\end{itemize}

\begin{figure}
\centering
\includegraphics[width=0.45\textwidth, clip]{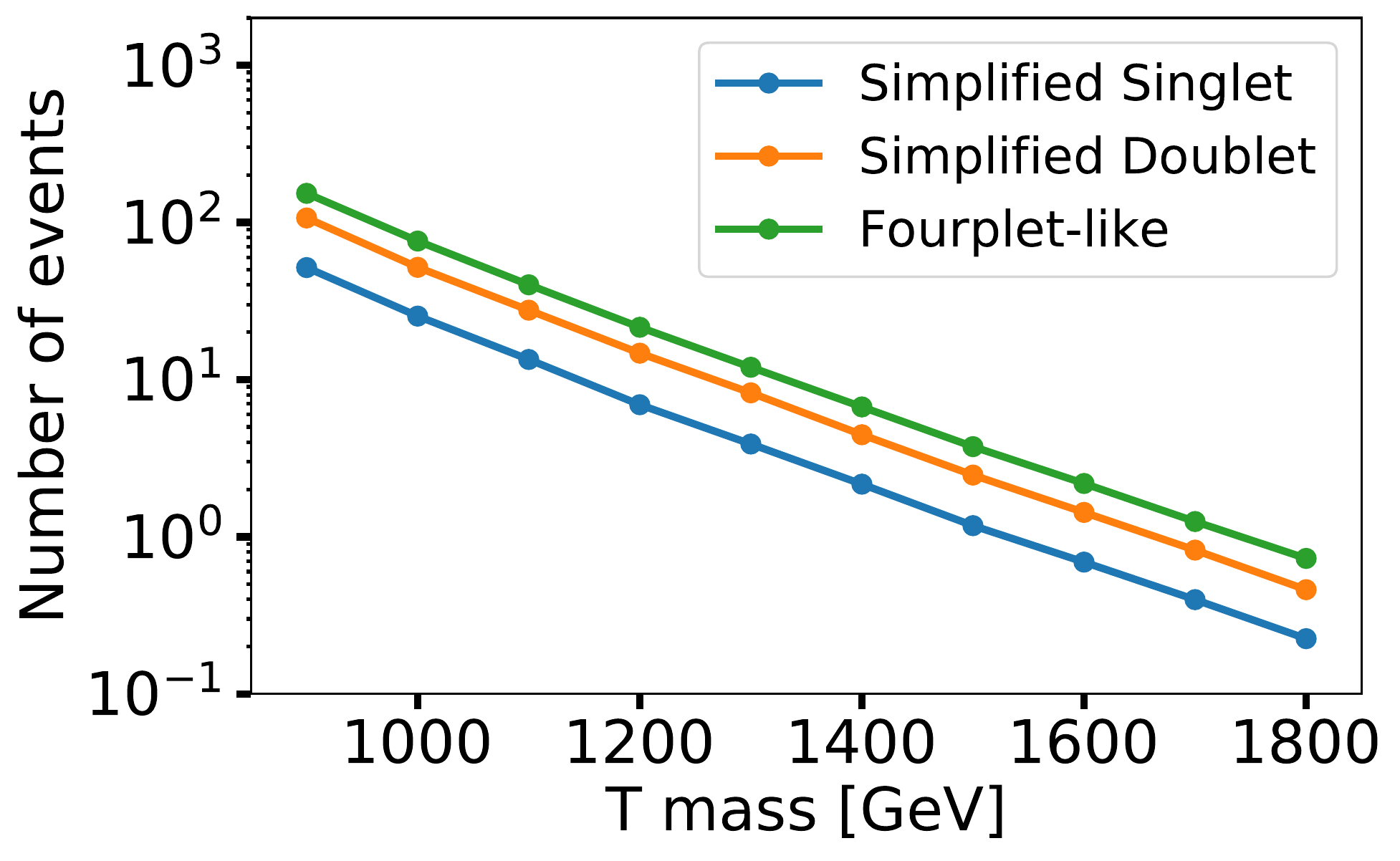}
\hspace{4mm}
\includegraphics[width=0.45\textwidth, clip]{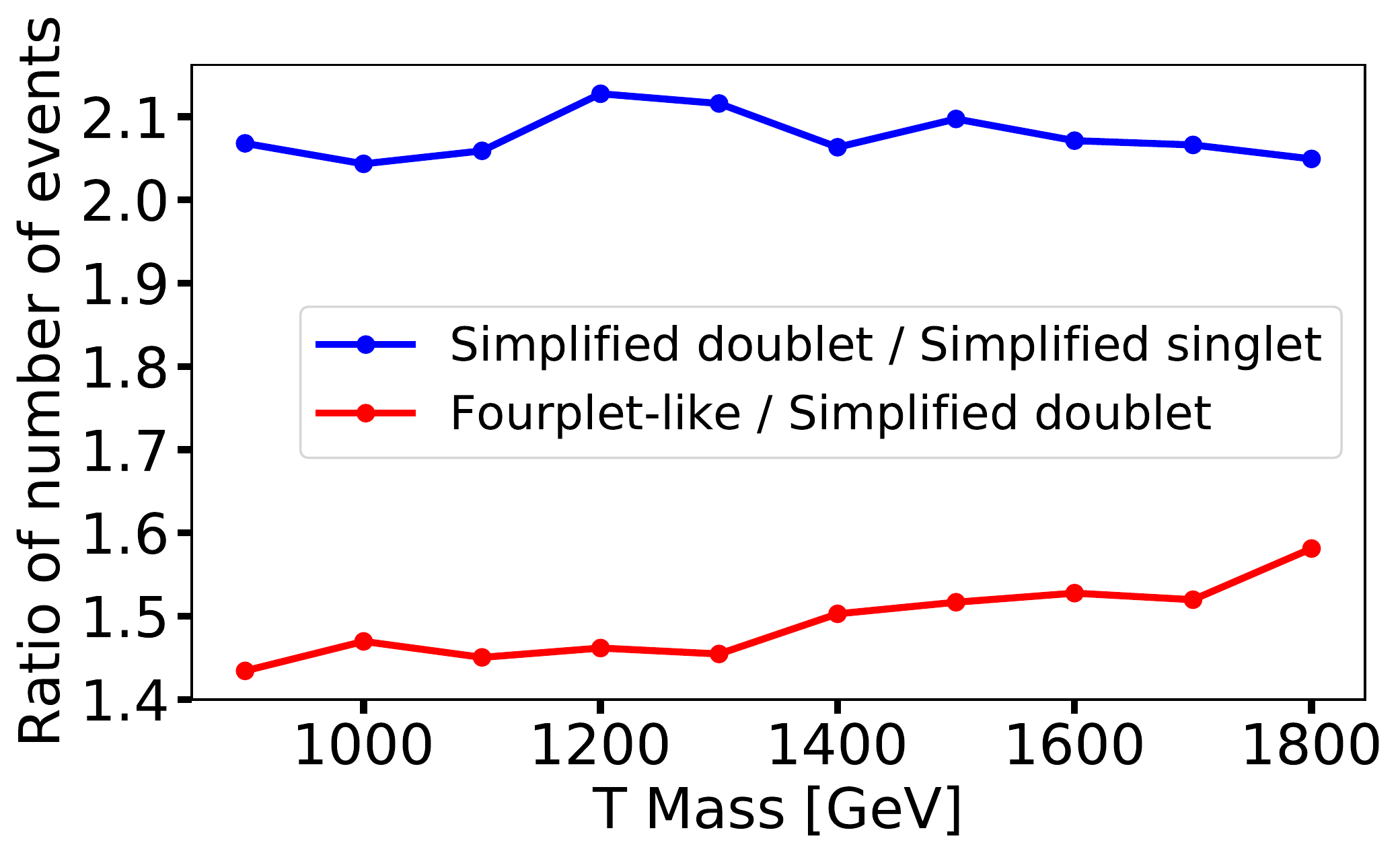}
\caption{(left) Number of events passing the 2SSL cuts in each scenario from table~\ref{T_scenarios}. (right) Ratio of the number of events between the fourplet-like and the simplified doublet scenarios (in red) and between the simplified doublet and the simplified singlet scenario (in blue). }
\label{fig:T_mass_scan}
\end{figure}

On the left of figure~\ref{fig:T_mass_scan} we show  the number of events passing the cuts for each $T$ mass in the different scenarios, on the right we show the ratio between those numbers. One expects that, to first order, an increase in the number of events will lead to a proportional decrease in the experimental upper limit, and we make that assumption here. We can check this assumption using the simplified doublet to simplified singlet ratio (blue curve on the right of figure~\ref{fig:T_mass_scan}). In the mass region analysed this ratio is around $2.1$, we can compare this with the ratio between the observed upper limits for these two scenarios in~\cite{tsearch}. Figure~\ref{fig:cms_results_T} only shows the limits for the combination of all channels, but considering the 2SSL channel alone that ratio is around $1.8$~\cite{private.T}, which is similar to the one we obtain. 

\begin{figure}
\centering
\includegraphics[width=0.6\textwidth]{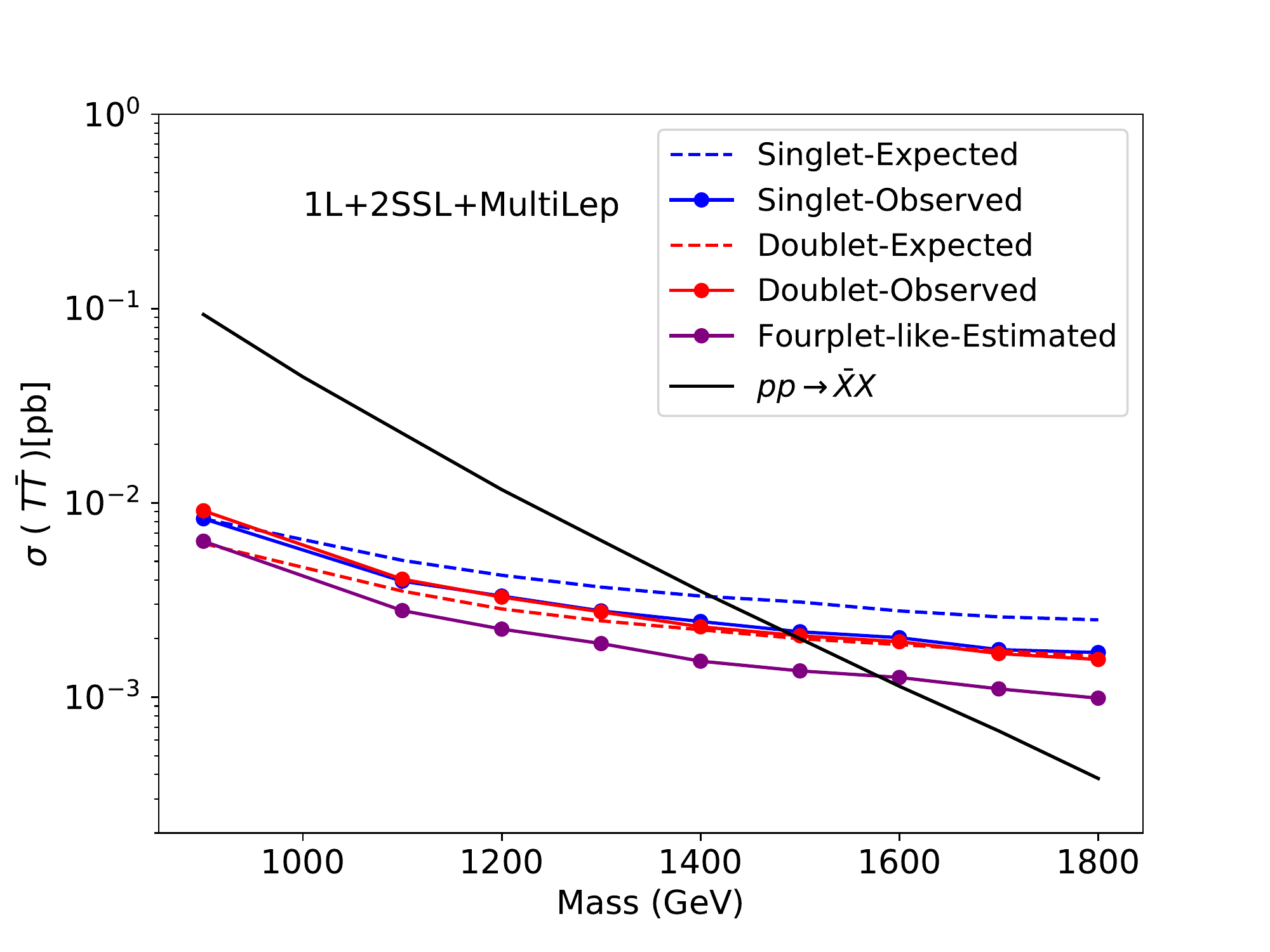}
\caption{Expected and observed upper limits taken from Fig.~\ref{fig:cms_results_T} for the simplified singlet (in blue) and simplified doublet (in red) scenarios. The purple line represents the estimated limits in the fourplet- like scenario; 1L+2SSL+MultiLep refers to the results from the 3 corresponding channels analysed in~\cite{tsearch}.}
\label{fig:T_limit_estimation}
\end{figure}

We can now focus on the comparison between the simplified doublet and fourplet-like scenarios. The red curve in figure~\ref{fig:T_mass_scan} shows a ratio around $1.5$ in the direction of increasing the number of events. The main effect here is that the presence of a 3-body decay into $WWt$ increases the probability of finding same sign leptons, even with a small branching ratio into that channel (we have $\text{BR}[W^+W^-t] = 0.1$). We expect the same effect to be present in all multi-lepton channels. We can now divide the simplified doublet upper limit by the ratio for each mass to estimate the upper limit of the fourplet-like scenario\footnote{Here we make another approximation, as the limits in figure~\ref{fig:cms_results_T} are for the three combined channels, and the ratios were obtained for the 2SSL channel alone.}, with results shown on figure~\ref{fig:T_limit_estimation}. In this very rough approximation the present exclusion would increase to $1.6$ TeV from the $1.5$ TeV obtained for the doublet in~\cite{tsearch}. Despite the roughness of this analysis, we firmly believe it motivates a new analysis by CMS that relaxes the assumption of 2-body decays only, as taking the 3-body decays into consideration will probably increase the mass exclusion using the same data available today (and such decays are present in most realistic MCHMs).

\subsection{Effect on the $\X$ search}
\label{Xrecast}

\begin{figure}
\centering
\includegraphics[width=0.8\textwidth]{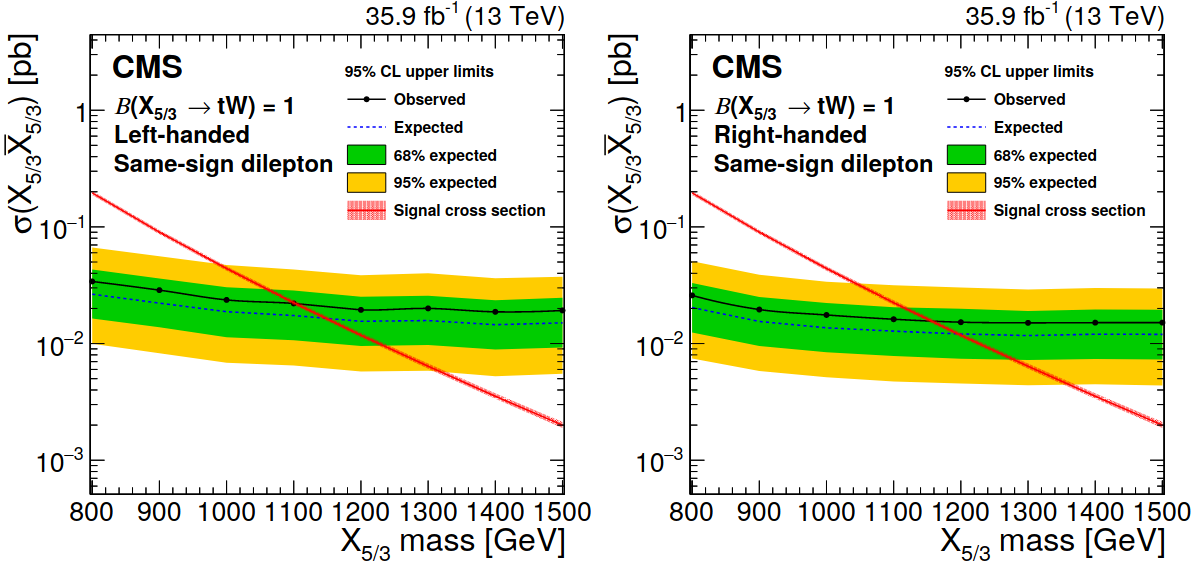}
\caption{Expected and observed upper limits of the signal cross section at 95$\%$ CL for an LH (left) and RH (right) $X_{5/3}$ from the same-sign dilepton search performed by~\cite{xsearch}. The band around the theoretical prediction shows the theoretical uncertainty. Figure extracted from~\cite{xsearch}.}
\label{fig:cms_results_x53}
\end{figure}

In the case of the $\X$ we follow very closely the strategy of the previous section, now using~\cite{xsearch} as the experimental search to be recast. In~\cite{xsearch} a single decay channel is considered, $\X \rightarrow W^+ t$, and the search is done separately for both pure left-handed and  right-handed $\X$. The cuts applied to the two chiralities are the same though, and the limits obtained are similar ($1.33$ TeV for $\X^{R}$ and $1.30$ TeV for $\X^{L}$), so we can expect to get a good estimate for the inclusion of 3-body decays in the vectorlike case treated here by using the same cuts. Two channels are analysed, the 2SSL and the single-lepton case, and results are presented for each channel and also for the combination. The results obtained for the 2SSL channel, which is the addressed channel here, can be seen in figure~\ref{fig:cms_results_x53}.

\begin{figure}
\centering
\includegraphics[width=0.3\textwidth]{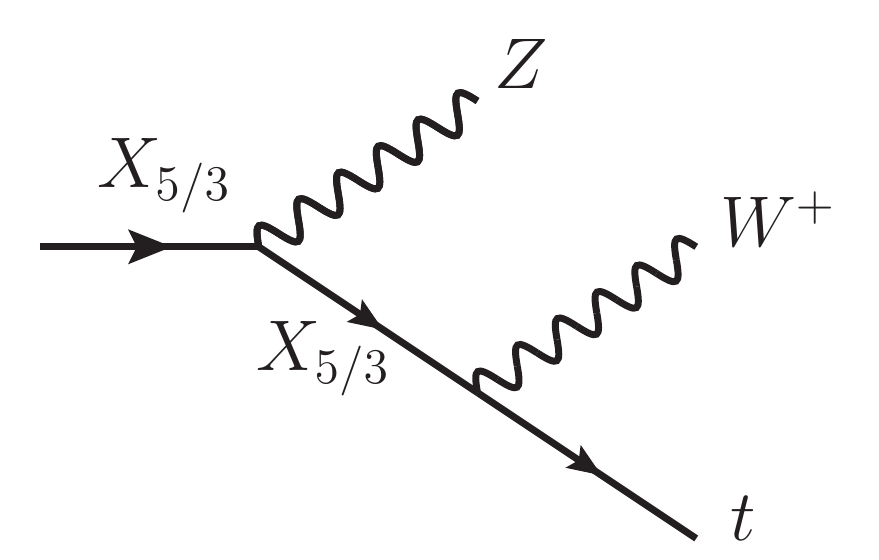}
\includegraphics[width=0.3\textwidth]{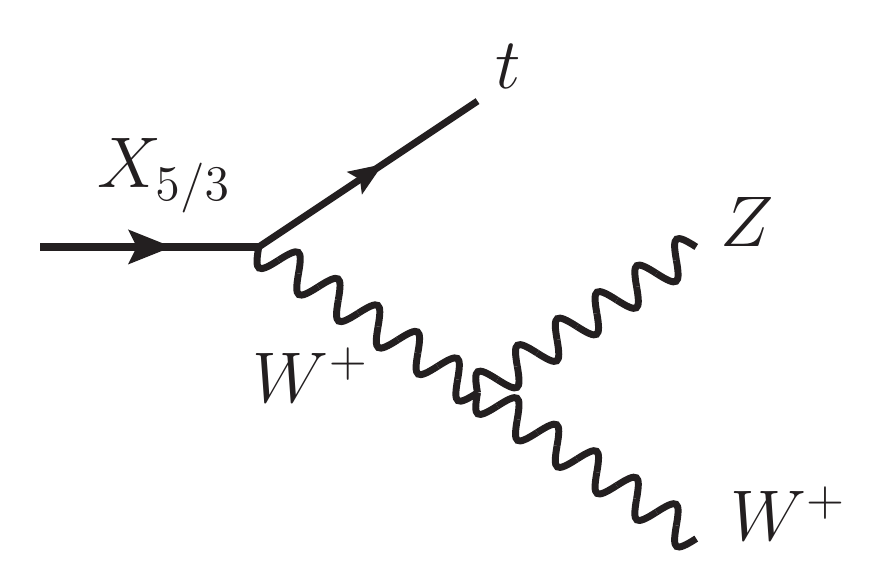}
\includegraphics[width=0.3\textwidth]{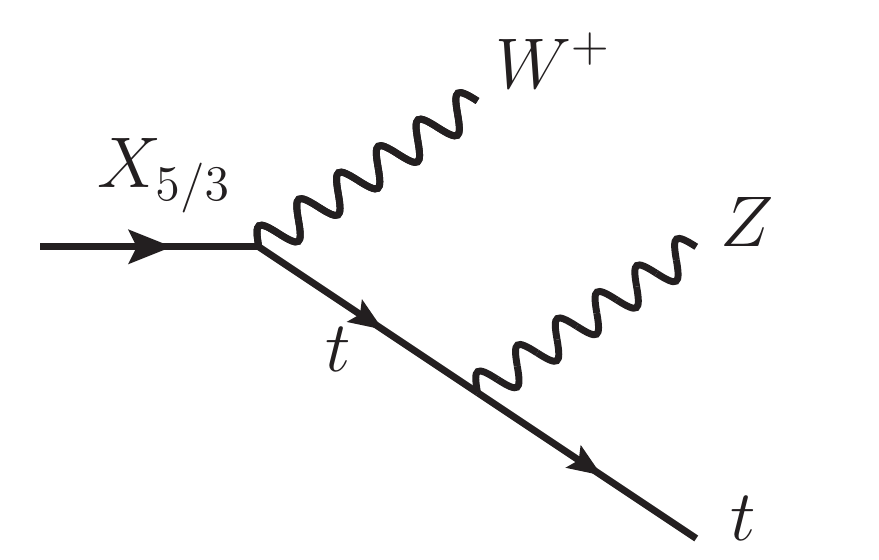}
\includegraphics[width=0.3\textwidth]{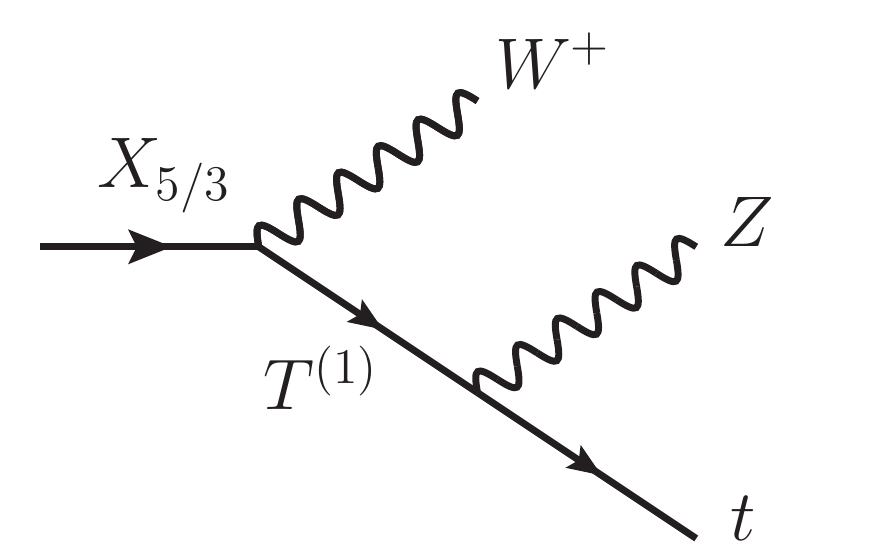}
\caption{Feynman diagrams of the three body decay $X_{5/3}\rightarrow W^+tZ$.  }
\label{fig:xTOwtz}
\end{figure}

\begin{figure}
\centering
\includegraphics[width=0.3\textwidth]{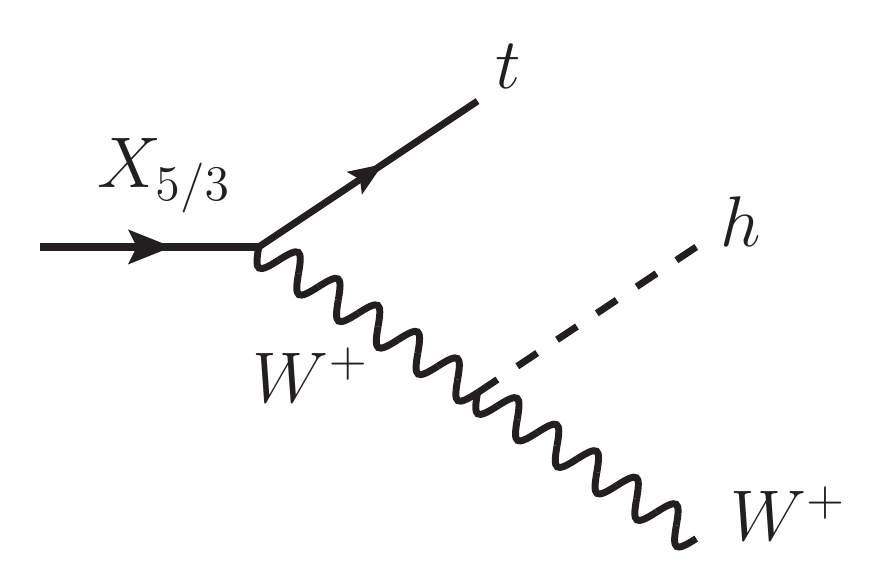}
\includegraphics[width=0.3\textwidth]{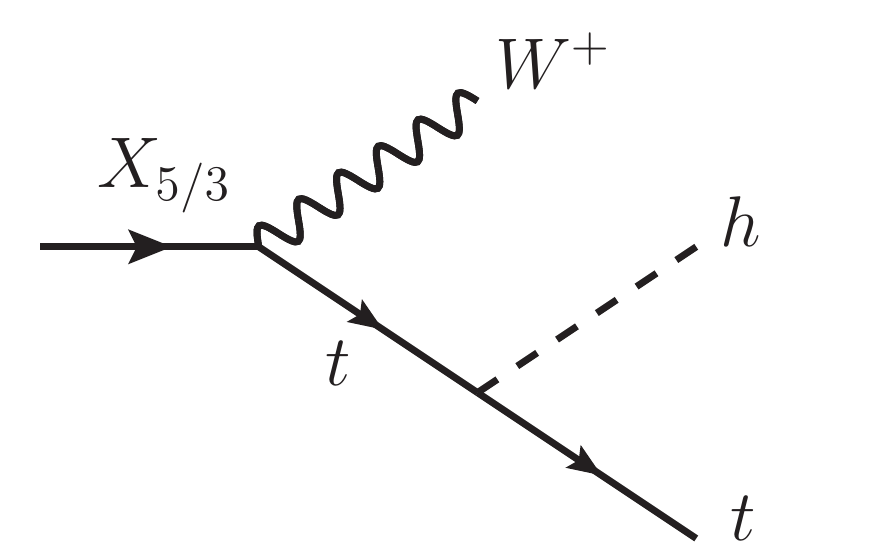}
\includegraphics[width=0.3\textwidth]{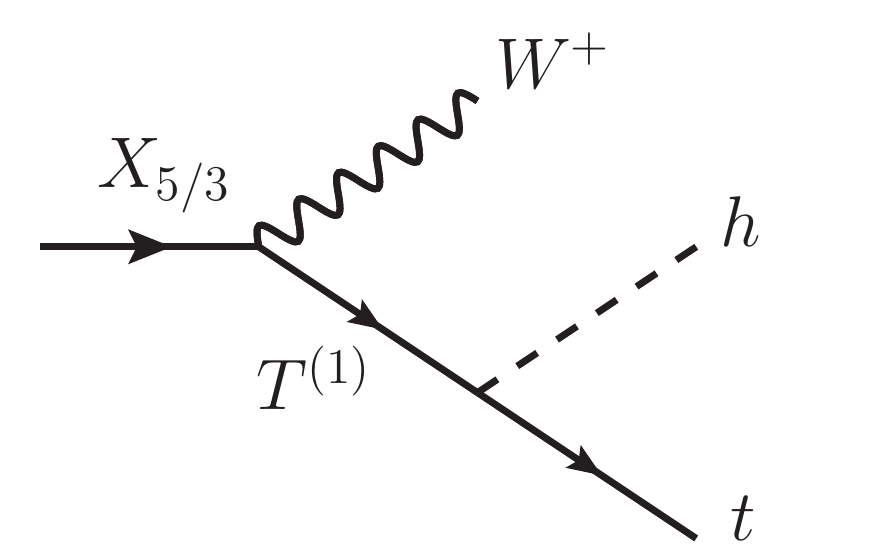}
\caption{Feynman diagrams of the three body decay $X_{5/3}\rightarrow W^+th$. }
\label{fig:xTOwth}
\end{figure}

Since we are focusing on the fourplet-like scenario, the main 3-body decays are $X_{5/3}\rightarrow W^+th$ and $X_{5/3}\rightarrow W^+tZ$, the relevant diagrams are shown in figures \ref{fig:xTOwtz} and \ref{fig:xTOwth}. In the majority of the parameter space points scanned the branching ratios in these two channels are similar, so we work with the 3-body scenario shown in table~\ref{x53_scenarios}. 
\begin{table}[hh]
  \centering
  \begin{tabular}{c|ccc}
     Scenario &  $\text{BR}(W^+t)$ & $\text{BR}(W^+tZ)$ & $\text{BR}(W^+th)$  \\
    \hline
    \textit{2-body}& $1$ & $0$ & $0$ \\
    \textit{3-body} & $0.8$ & $0.1$ & $0.1$ \\
    \end{tabular}
    \caption{Scenarios considered in the $X_{5/3}$ analysis and their corresponding branching ratio configurations.}
    \label{x53_scenarios}
\end{table}

\begin{figure}
\centering
\includegraphics[width=0.6\textwidth]{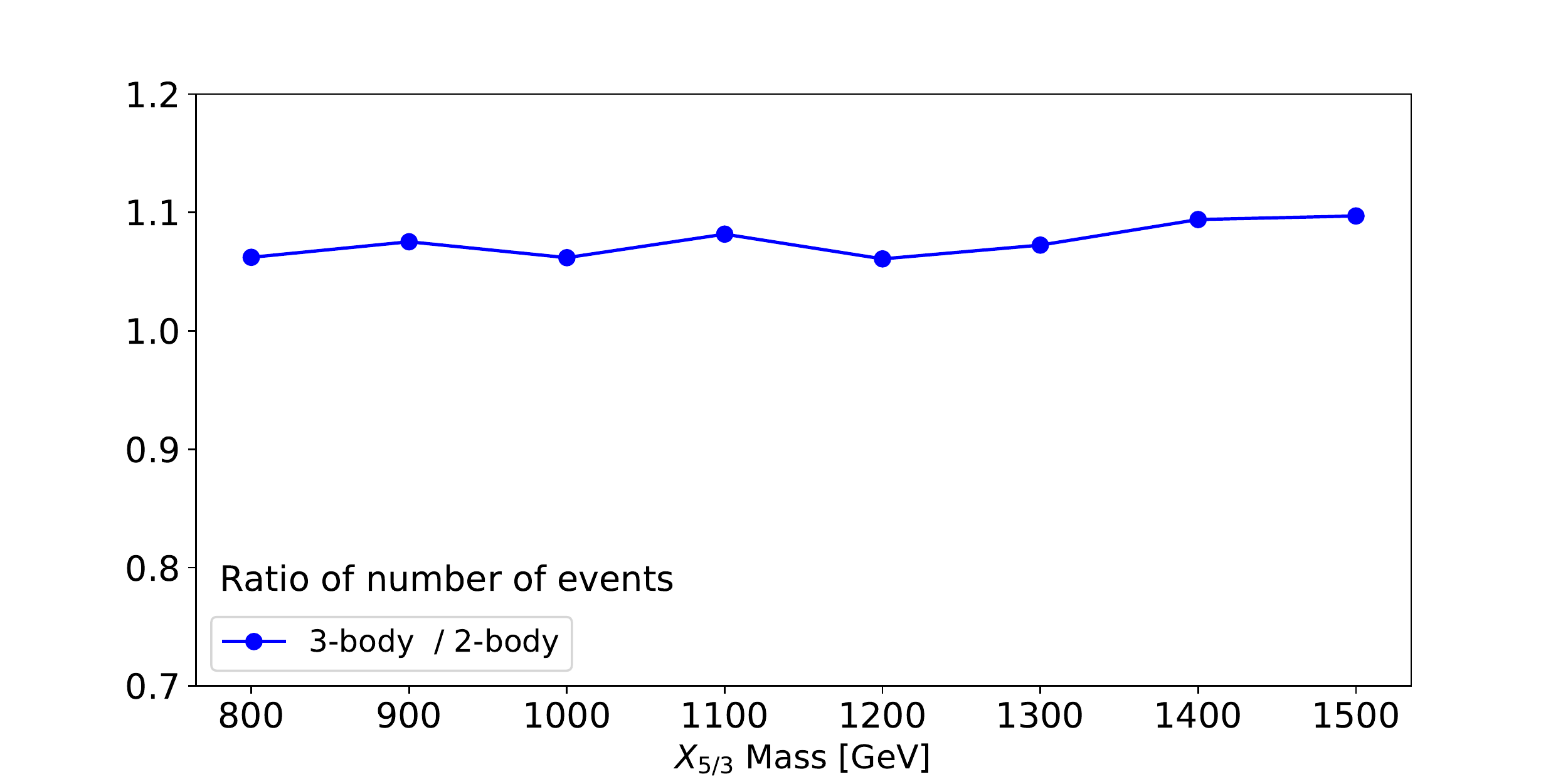}
\caption{ $X_{5/3}$ resonances search: ratio of the number of events passing the kinematical cuts in the 2-body and the 3-body scenarios.}
\label{fig:x53_mass_scan_events_ratios}
\end{figure}

The simulation and cut flow follows closely what was done in section~\ref{T1recast}, with the following changes: the subleading lepton is required to have $p_T>35\text{ GeV}$, number of jets $N_j \geq 2$, the total number of constituents\footnote{The number of constituents is equal to number of jets plus number of leptons beyond the two considered for the lepton pair} $N_{\text{const}}\geq5$ and $H^{\text{lep}}_T>1200 \text{ GeV}$. The ratio between the number of events passing the cuts in each scenario can be seen in figure~\ref{fig:x53_mass_scan_events_ratios}. 
The effect of the 3-body decays goes in the same direction but is clearly smaller than in the $\Tone$ case, with an increase in the number of events between 5\% and 10\%. This is a consequence of the fact that the 3-body decays now introduce extra $Z$ or $h$, instead of extra $W$'s, and those do not contribute as strongly to the same sign dilepton channel. The effect is correspondingly smaller in the recasting to the mass exclusion limit, so no significant change to the limit is obtained.

\section{Inclusive search for vectorlike resonances in the presence of 3-body decays}
\label{searchstrat}

Now, motivated by the fact that including 3-body decays can increase the experimental sensitivity to VLQ, we also relax the very common assumption that there is only one light VLQ. The simplest scenario in a complete model that provides us two light VLQs comes from the \five with a fourplet-like $\Tone$. In~\cite{Bautista:2020mxw} we have found a few benchmark points which are good representatives of this situation and we will use one of those, named $C_9$ in~\cite{Bautista:2020mxw}, to propose a search strategy. The main phenomenological characteristics of the model can be seen in table~\ref{fig:C9_properties}. This point was chosen because it reproduces well the fourplet-like scenario of the previous chapter (specifically the 3-body decay branching ratios) and the masses of the low lying resonances are close to experimental limits\footnote{
The updated search in~\cite{tsearch} was published in the final stages of this work, after this analysis was finished, increasing the $2 \sigma$ constraint to $m_{\Tone} \approx 1.5$ TeV from previous constraints lying around $1.3$ TeV to $1.4$  TeV (depending of decay assumptions)~\cite{CMS:2017ynm,CMS:2018zkf,CMS:2018wpl,CMS:2019eqb,CMS:2020ttz,ATLAS:2018ziw}. The results of this section can be easily extrapolated for small increases in mass. 
}.

\begin{table}
  \centering
  \begin{tabular}{cccccc}
    \cline{1-6}
     &  $T^{(1)}$ &  $T^{(2)}$ &  $T^{(3)}$ &  $B$ &  $X_{5/3}$ \T\B \\
    \hline
    Mass (TeV) & 1.3 & 1.8 & 2.0 & 2.0 & 1.3 \TB \\
    \hline
    Width (GeV) & 7.8 & 13.4 & 6.8 & 5.5 & 6.7 \TB \\
    \hline
    Pair production $\sigma$ (fb) & 6.6 & 0.50 & 0.17 & 0.21 & 6.7 \TB \\
    \hline
    BR($th$)& 0.46 & 0.16 & 0.03 & - & - \T\\
    BR($tZ$) & 0.39 & 0.07 & 0.14 & - & -  \\
    BR($W^+b$) & 0.02 & 0.20 & 0.14 & - & - \\
    BR($W^-t$)& - & - & - & 0.05 & - \\
    BR($W^+t$) & - & - & - & - & 0.86  \\
    BR($W^+W^-t$) & 0.10 & 0.12 & 0.01 & - & - \\
    BR($W^+tZ$)& - & - & - & - & 0.03 \\
    BR($W^+ht$) & - & - & - & - & 0.03  \\
    BR($X_{5/3}W^-$) & - & 0.13 & 0.01 & - & -  \\
    BR($T^{(1)}h$) & - & 0.07 & 0.01 & - & - \\
    BR($T^{(1)}Z$) & - & 0.06 & 0.01 & - & - \\
    BR($T^{(2)}h$) & - & - & 0.18 & - & - \\
    BR($T^{(2)}Z$) & - & - & 0.42 & - & - \\
    BR($W^-T^{(2)}$) & - & - & - & 0.77 & - \\
    Other BRs & 0.03 & 0.19 & 0.05 & 0.18 & 0.08 \B \\
    \hline
    \end{tabular}
    \caption{Masses, decay widths and branching ratios of the resonances in the benchmark point $C_9$.}
    \label{fig:C9_properties}
\end{table}

\begin{figure}
\centering
\begin{subfigure}[b]{0.45\textwidth}
    \centering
    \includegraphics[width=0.85\textwidth]{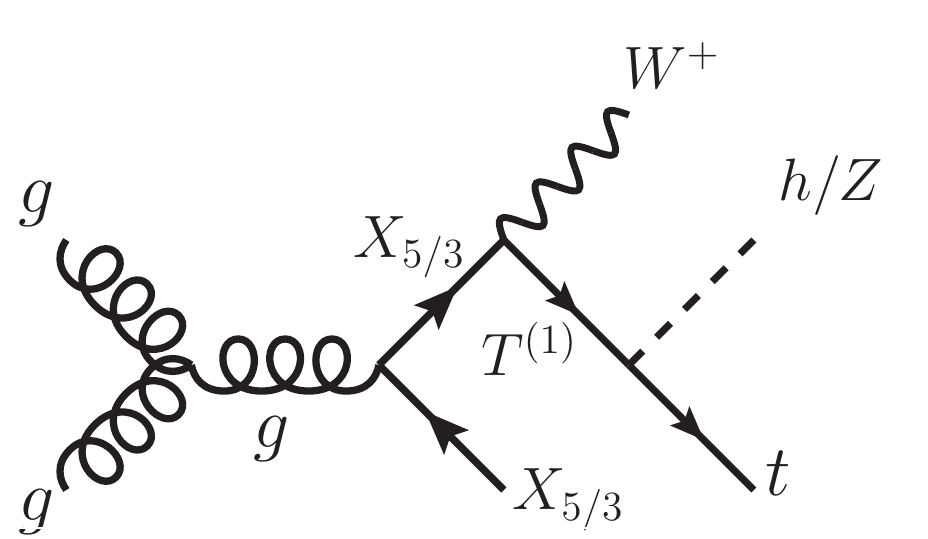}
    \subcaption{}\label{fig:x53_TO_wT_TO_thZ}
\end{subfigure}
\begin{subfigure}[b]{0.45\textwidth}
    \centering
    \includegraphics[width=0.85\textwidth]{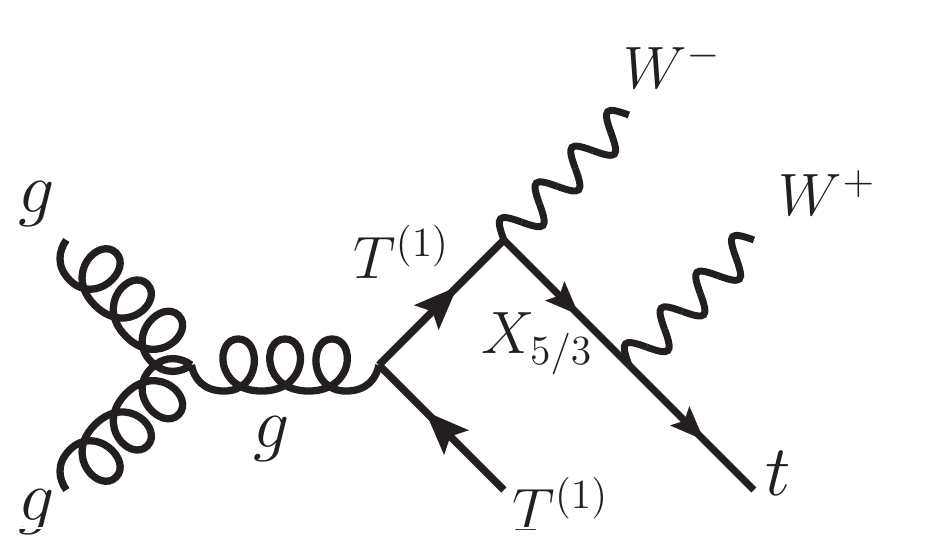}
    \subcaption{}\label{fig:T_TO_wx53_TO_wt}
\end{subfigure}
\caption{Feynman diagrams of the processes involving both two- and tree-body decays of resonances.}
\label{fig:double_prod_3body_decay}
\end{figure}

The key characteristic of a model with a fourplet-like $\Tone$ is the degeneracy in mass with the $\X$. Taken together with the presence of 3-body decays, this makes it non-trivial to separate searches for these two resonances. To clarify this point, consider the two diagrams in figure~\ref{fig:double_prod_3body_decay}. Due to their degenerate masses the $\X$ and $\Tone$ cannot be both on-shell in the upper decay chain in those diagrams. This means that, for instance, diagram~\ref{fig:x53_TO_wT_TO_thZ} can generate two different events: (i) the production of an on-shell $\X$, followed by its 3-body decay; or (ii) the $\X$-mediated production of on-shell $\Tone + W$, followed by a 2-body decay of $\Tone$. The key point here is that the final states for these two situations are the same (an analogous situation occurs in diagram~\ref{fig:T_TO_wx53_TO_wt}, with the roles of $\Tone$ and $\X$ reversed). This also happens for other combinations of 2-body and 3-body decays, but the decays in figure~\ref{fig:double_prod_3body_decay} are the dominant 3-body decays in the fourplet-like scenario.

\begin{figure}
\centering
\begin{subfigure}{0.49\textwidth}
\centering
\includegraphics[width=1\textwidth]{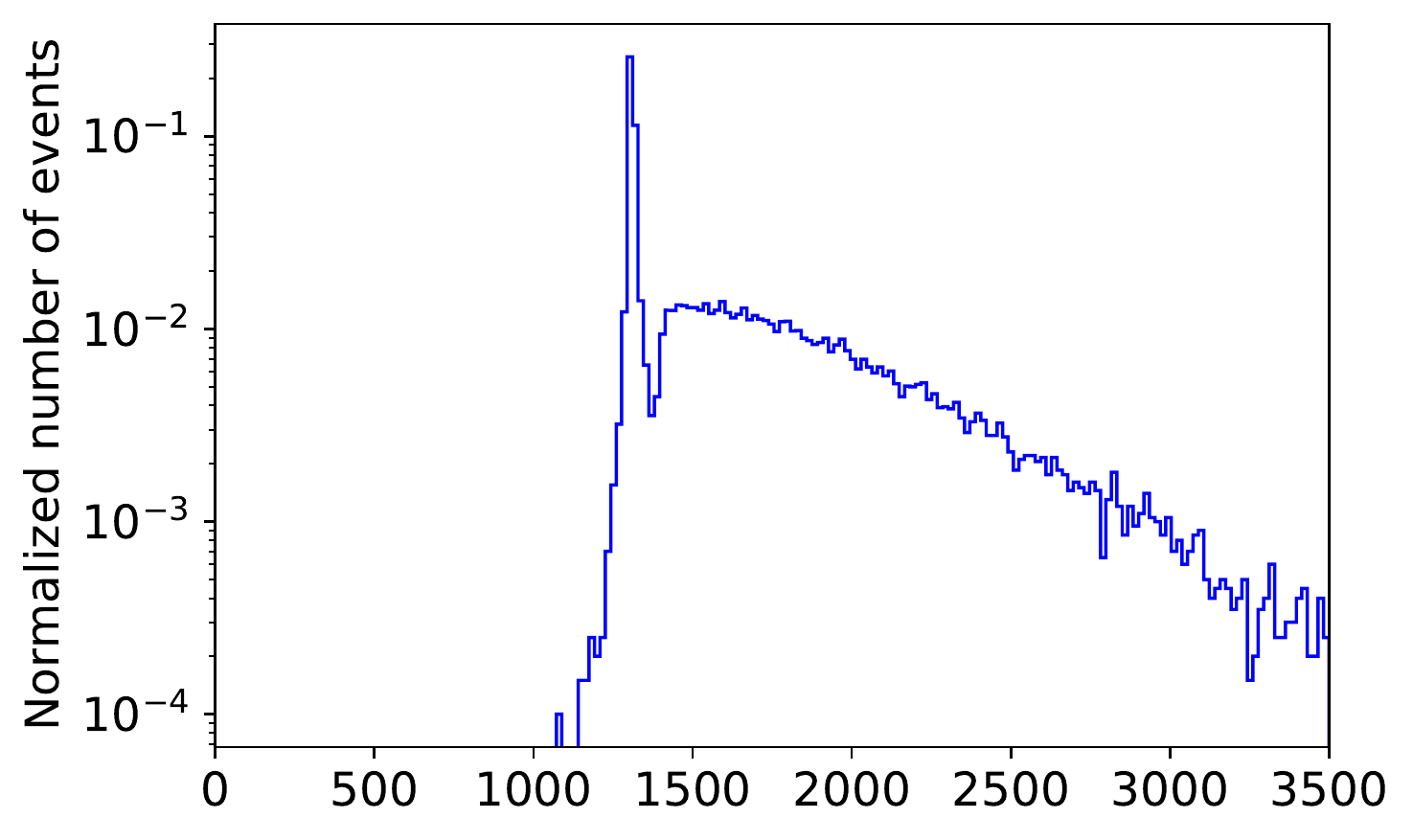}
\subcaption{$\text{M}[W^+,t,Z]$ (GeV)} \label{fig:distr_x53_wtz_fourplet}
\end{subfigure}
\begin{subfigure}{0.49\textwidth}
\centering
\includegraphics[width=1\textwidth]{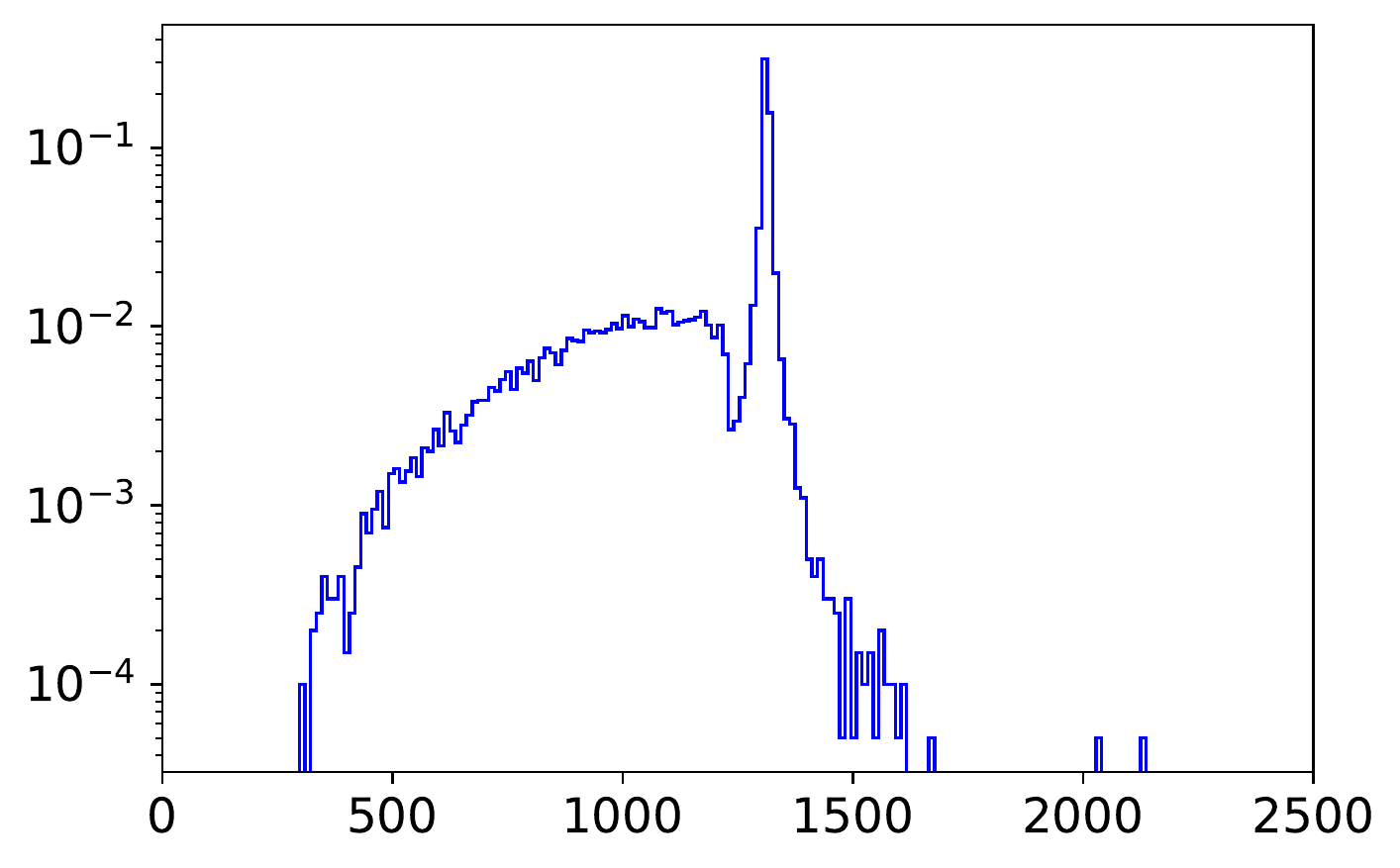}
\subcaption{$\text{M}[t,Z]$ (GeV)} \label{fig:distr_x53_tz_fourplet}
\end{subfigure}
\caption{Invariant mass distributions of the process in Figure~\ref{fig:x53_TO_wT_TO_thZ} for the benchmark point $C_9$ (fourplet-like point). Histograms generated with MadAnalysis~5~\cite{Conte:2012fm}. }
\label{fig:distr_x53_fourplet}
\end{figure}

Figure~\ref{fig:distr_x53_fourplet} shows how these two contributions can mix in kinematic variables, in~\ref{fig:distr_x53_tz_fourplet} one can clearly see two features: (i) the peak generated at $1.3$ TeV generated by decays of on-shell $\Tone$; and (ii) the bump in the region $\text{M}[t,Z] \lesssim 1.3$ generated by 3-body decays off on-shell $\X$  (which force $\Tone$ off-shell). This illustrates the difficulty in separating the two signals: in the example a lot of $t Z$ events will be coming from off-shell $\Tone$ even when the resonance itself is narrow, which is a counter-intuitive result. Even if the peak is properly identified, conclusions about the production cross-section will be affected by the fact that it is sitting on top of another new physics signal. The same applies to the invariant mass $\text{M}[W^+,t,Z]$, in~\ref{fig:distr_x53_wtz_fourplet}, where we see a peak from on-shell $\X$ sitting on top of a off-shell $\X$ bump. 

It is important to realize that this is a general feature of the degeneracy in mass and the presence of 3-body decays, and will be present in any model as long as the involved coupling constants are sizeable. In the \five this is guaranteed by the fact that when the $\Tone$ and the $\X$ are coming from the same multiplet (i.e. we have a fourplet-like $\Tone$), the couplings will be significant and the states degenerate, which implies the 3-body decays will be relevant.
 
\subsection{Signal}
\label{subsec.signal}

Instead of going through the extra problem of trying to disentangle these two states, we here propose the alternate strategy of using this in our favor. By looking for new physics in an inclusive way, considering contributions of pair production of both the $\Tone$ and the $\X$ to the same channel, we will have increased sensitivity to the new physics. The dominant 3-body decays of the $\X$ are into $W^+ t h$ and $W^+ t Z$. We can safely neglect the case where both $\X$ in the pair decay into three bodies, as the branching ratio is too small, so the dominant decay in the other leg will be into $W^+ t$. The same can be obtained from the pair production of fourplet-like $\Tone$, as the dominant 3-body decay is $W^+ W^- t$ and the 2-body decays are dominantly $t h$ and $t Z$ (again we can neglect two 3-body decays). There is also a contribution from heavier top partners.
Although the cross section of the $\Ttwo$ pair production is around 10\% of the $\Tone$ pair production, the 2-body channels considered have lower branching ratios. In the case of the $W^+ W^- t$ channel, although the $\Ttwo$ branching ratio is a little higher, it does not get off shell contributions as big as those of the $\Tone$. 
Therefore the actual contribution of the $\Ttwo$ resonance is around 2\% of the events generated and $\Tthree$ is even smaller. Hence, results found here should also apply well to models where other top partners are not present.

We start thus from $t \bar{t}\, W^+ W^-$ and a $h$ or $Z$ and, to maximize the number of events, we choose the $b \bar{b}$ decay for the Higgs or the $Z$ bosons. Considering the decays of the tops, our signal becomes $W^+W^-W^+W^-b\bar{b}b\bar{b}$ ($4W4b$). The presence of the four W bosons allows us to explore multi-leptonic channels, and we will focus in two channels: one containing two leptons (meaning electrons or muons) of the same charge (2SSL) and the other containing 3 leptons irrespective of charge (3L). We also consider the 2SSL+3L channel, which includes events selected for either of the previous channels. Since it would be unrealistic to demand the full reconstruction of four $W$ and four $b$, we will also have new physics contributions coming from 2-body decays only (which can imply less $b$ quarks after decays), and we include those too. The processes contributing to the signal are listed in table~\ref{fig:signalTable}. A noteworthy feature is that the cross sections in the second column of table~\ref{fig:signalTable} can not be consistently estimated by the product of the corresponding pair production cross sections and the branching ratios into 2- and 3-body channels listed in table~\ref{fig:C9_properties}, despite the fact that the VLQs are narrow. Such an estimation works in the case of 2-body decays only. In the case of 3-body decays the cross sections are around four times bigger than expected, and this is a direct consequence of the off-shell contributions discussed previously (the amount of ``off-peak'' events in figure~\ref{fig:distr_x53_fourplet} makes that clear). This is one of the advantages of this inclusive search.

\begin{table}
  \centering
  \begin{tabular}{|c|c|c|c|}
    \cline{1-4}
    Process & $\sigma$ [fb] &   decay mode &  $\sigma \times \text{BR}$ [ab] \\
    \hline
    $X_{5/3} \bar{X}_{5/3} \rightarrow \ttWW$ 
          & 4.87 & \Wleppm\Wleppm\Whad\Whad & 208 \\  
          &      & \Wleppm\Wlepmp\Wleppm\Whad & 133 \\
          &      & \Wleppm\Wlepmp\Wleppm\Wlepmp & 106 \\
    \hline
    $X_{5/3} \bar{X}_{5/3} \rightarrow$ \WWttH
          & 1.12 & \Wleppm\Wleppm\Whad\Whad & 27.6 \\
          &      & \Wleppm\Wlepmp\Wleppm\Whad & 17.7 \\
          &      & \Wleppm\Wlepmp\Wleppm\Wlepmp & 1.41 \\
    \hline
    $T\bar{T} \rightarrow$ \WWttH 
          & 1.01 & \Wleppm\Wleppm\Whad\Whad & 24.9 \\
          &      & \Wleppm\Wlepmp\Wleppm\Whad & 15.9\\
          &      & \Wleppm\Wlepmp\Wleppm\Wlepmp & 1.27 \\
    \hline
    $T\bar{T} \rightarrow$ \ttHH
          & 1.37 & ($hh \rightarrow b\bar{b} W^+ W^-$) &  \\
          &      & \Wleppm\Wleppm\Whad\Whad & 14.6 \\
          &      & \Wleppm\Wlepmp\Wleppm\Whad & 9.32 \\
          &      & \Wleppm\Wlepmp\Wleppm\Wlepmp & 0.75 \\
    \hline
    $X_{5/3} \bar{X}_{5/3} \rightarrow$ \WWttZ
          & 1.1 & \Wleppm\Wleppm\Whad\Whad \Zb & 7.15 \\
          &      & \Wleppm\Wlepmp\Wleppm\Whad \Zb& 4.58 \\
          &      & \Wleppm\Wlepmp\Wleppm\Wlepmp \Zb & 3.66 \\
    \hline
    $T\bar{T} \rightarrow$ \WWttZ
          & 0.86 & \Wleppm\Wleppm\Whad\Whad \Zb& 5.59 \\
          &      & \Wleppm\Wlepmp\Wleppm\Whad\Zb & 3.57 \\
          &      & \Wleppm\Wlepmp\Wleppm\Wlepmp\Zb & 0.29 \\
    \hline
    $T\bar{T} \rightarrow$ \ttZH
          &      & \Wleppm\Wlepmp\Zlep & 4.29 \\ \cline{3-4}
          &      & ($h \rightarrow W^- W^+$)  &   \\
          &      & \Wleppm\Wleppm\Whad\Whad\Zb & 3.33 \\
          &      & \Wleppm\Wlepmp\Wleppm\Whad\Zb & 2.13 \\
          &      & \Wleppm\Wlepmp\Wleppm\Wlepmp\Zb & 0.17 \\
    \hline
    $T\bar{T} \rightarrow$ \ttZZ
          & 1.03 & \Wleppm\Wlepmp\Zlep \Zb & 0.98 \\
    \hline
    \end{tabular}
    \caption{
        Signal processes for the point $C_9$ in the 2SSL search at LO and $\sqrt{s}=14$ TeV. Here T stands for $\Tone$, $\Ttwo$ or $\Tthree$. The second column indicates cross section before decays, the third indicates the decay mode of the vector bosons (with $V_l$, $V_\text{had}$ and $V_{b \bar{b}}$ meaning decays into leptons, hadrons and $b \bar{b}$  respectively) and the forth is the cross section after the indicated decay (with $t \rightarrow b W$ and $h \rightarrow b \bar{b}$ where not otherwise indicated). The cross sections were computed at LO and $\sqrt{s}=14$ TeV.
    }
    \label{fig:signalTable}
\end{table}

\subsection{Backgrounds}
\label{subsec.back}

On the subject of backgrounds, the first important observation is that the $4W4b$ signal is also generated by four top production in the SM.  The $t\bar{t}t\bar{t}$ signal has been intensively searched for~\cite{Alvarez:2016nrz,CMS:2019rvj,ATLAS:2020hpj} and we can profit from the accumulated background knowledge, since we will have the same backgrounds present in those searches. Those backgrounds are shown in table~\ref{fig:backgroundTable2SSL}, where one can see that main irreducible backgrounds to the 2SSL and 3L channels come from the production of $t \bar{t}$ pair in association with a boson and $tZbjj$. These are followed in importance by the production of a top pair plus two bosons, but we will neglect the case were those bosons are a $Z$ or a $h$ as in both cases the decay into leptons is small when compared with the $W$, so only $\ttWW$ is included. Of course $t\bar{t}t\bar{t}$ itself is a background in our case.

\begin{table}
  \centering
  \begin{tabular}{|c|c|c|c|}
    \cline{1-4}
    Backgrounds & $\sigma$ [fb] &   decay mode &  $\sigma \times \text{BR}$ [fb]\\
    \hline
    \ttW + jets & 574.5 & \Wleppm\Wleppm\Whad & 18.14 \\
                &       & \Wleppm\Wleppm\Wlepmp & 5.81 \\
    \hline
    \ttZ + jets & 743.1    & \Wleppm\Whad\Zlep & 12.73 \\
                &       & \Wleppm\Wlepmp\Zlep & 2.04\\
    \hline
    \ttH & 479.9 & ($h \rightarrow W^- W^+$) & \\
         &       & \Wleppm\Whad\Wleppm\Whad & 4.42 \\
         &       & \Wleppm\Wlepmp\Wleppm\Whad & 2.82 \\
         &       & \Wleppm\Whad\Zlep\Zhad & 0.54 \\
         &       & \Wleppm\Wlepmp\Wleppm\Wlepmp & 0.22    \\
    \hline
    $tZbjj$ & 317 & \Wleppm\Zlep & 4.6  \\
    \hline
    \tttt & 11.8 & \Wleppm\Wleppm\Whad\Whad & 0.51 \\
          &      & \Wleppm\Wlepmp\Wleppm\Whad & 0.32 \\
          &      & \Wleppm\Wlepmp\Wleppm\Wlepmp & 0.03 \\
    \hline
    \ttWW & 9.88 & \Wleppm\Whad\Wleppm\Whad & 0.42  \\
          &      & \Wleppm\Wlepmp\Wleppm\Whad & 0.27 \\
          &      & \Wleppm\Wlepmp\Wleppm\Wlepmp & 0.03 \\
    \hline
    \end{tabular}
    \caption{
       Most important background processes contributing to the 2SSL, 3L or 2SSL+3L channels, following the same conventions of table~\ref{fig:signalTable}. Here ``+jets'' refers to $0$, $1$ or $2$ jets generated at the hadronization stage of the simulation and $j$ stands for a hard light jet, generated at parton level simulation. The cross sections were computed at LO and $\sqrt{s}=14$ TeV.
    }
    \label{fig:backgroundTable2SSL}
\end{table}

\subsection{Proposed Search Strategy}
\label{subsec.strat}

All the cross-sections in tables~\ref{fig:signalTable} and~\ref{fig:backgroundTable2SSL} were obtained by simulation with Madgraph5 (v2.9)~\cite{Alwall:2014hca} at LO. The event samples are then passed through showering and hadronization, performed by Pythia8~\cite{SJOSTRAND2015159} (with jet matching done in the MLM
matching scheme), and through Delphes3 (v3.5.0)~\cite{de_Favereau_2014} for a fast detector analysis. In Delphes the default card for the HL-LHC was used and jet clustering is done by FastJet~\cite{CACCIARI200657,Cacciari_2012} using the anti-$k_T$ algorithm with $R = 0.4$. The $b$-jet reconstruction is done with an efficiency of $0.75(1-\frac{p_T}{5000\text{ GeV}})$. Leptons are isolated if the $p_T$ sum of all the particles inside the cone with fixed radius $\mathcal{R}=0.3$ around the lepton, divided by $p_T$ of the lepton, is less than 0.1. We also require that all reconstructed particles (leptons and jets) have $p_T>30$ GeV and $|\eta|<3$. The number of events obtained at $\mathcal{L}=4\text{ ab}^{-1}$ for signal and backgrounds after this minimal set of requirements can be seen in the ``No cuts'' column of tables~\ref{fig:events_results_2ssl},~\ref{fig:events_results_3l} and~\ref{fig:events_results_2ssl_3l} (respectively for the 2SSL, 3L, and 2SSL+3L signals).

\begin{table}
  \centering
  \begin{tabular}{|c|c|c|c|c|c|c|}
    \hline
    \multicolumn{3}{|c|}{\multirow{4}{*}{Process}} & \multicolumn{4}{c|}{Number of events - 2SSL ($\mathcal{L}=4\text{ ab}^{-1}$)}\\
    \cline{4-7}
    \multicolumn{3}{|c|}{\multirow{3}{*}{}} & \vtop{\hbox{\strut No cuts}} & \vtop{\hbox{\strut  $N_b\geq3$}}
    & \vtop{\hbox{\strut  $N_b\geq3$,}\hbox{\strut $H^{\text{lep}}_T>1.8$ TeV}}
    & \vtop{\hbox{\strut  $N_b\geq3$,}\hbox{\strut $H^{\text{lep}}_T>1.8$ TeV}\hbox{\strut $\cancel{E}_T>150$ GeV}}
    
    \\
    \hline
    \multirow{8}{*}{\rotatebox{90}{Signal}} & \multirow{4}{*}{\rotatebox[origin=c]{90}{2-bodies}} & $X\bar{X} \rightarrow \ttWW$ & $356$ & $40$ & $20$ & $15$ \T\\
    & & $T\bar{T} \rightarrow$ \ttHH& $39$ & $14$ & $5$ & $3$ \\
    & & $T\bar{T} \rightarrow$ \ttZH & $17$ & $7$ & $4$ & $2$\\
    & & $T\bar{T} \rightarrow$ \ttZZ& $12$ & $2$ & $1$ & $1$ \B\\
    \cline{2-7}
    & \multirow{4}{*}{\rotatebox[origin=c]{90}{3-bodies}} & $X\bar{X} \rightarrow$ \WWttH& $48$ & $24$ & $15$ & $11$ \T \\
    & & $T\bar{T} \rightarrow$ \WWttH & $42$ & $21$ & $13$ & $9$ \\
    & & $T\bar{T} \rightarrow$ \WWttZ& $15$ & $6$ & $4$ & $3$ \\
    & & $X\bar{X} \rightarrow$ \WWttZ& $12$ & $5$ & $3$ & $2$ \B \\
    \hline
    \multirow{6}{*}{\rotatebox{90}{Background}}&\multicolumn{2}{c|}{\ttW} & $20536$ & $691$ & $19$ & $9$ \T\\
    & \multicolumn{2}{c|}{\ttZ} & $7062$ & $237$ & $4$ & $2$ \\
    & \multicolumn{2}{c|}{\ttH} & $3893$ & $132$ & $1$ & $0$ \\
    & \multicolumn{2}{c|}{\tttt} & $658$ & $288$ & $6$ & $3$ \\    
    & \multicolumn{2}{c|}{\ttWW} & $597$ & $30$ & $1$ & $0$ \\    
    & \multicolumn{2}{c|}{tZ bjj} & $761$ & $18$ & $0$ & $0$ \B \\
    \hline
    \multicolumn{3}{|c|}{$S/B$} & & $0.1$ & $2.2$& $3.2$ \T\B \\
    \hline
    \multicolumn{3}{|c|}{$S/\sqrt{B}$} & & $3.2$ & $11.9$ & $12.1$ \T\B  \\
    \hline
    \end{tabular}
    \caption{Number of events surviving the cuts implementation in the 2SSL search channel.}
    \label{fig:events_results_2ssl}
\end{table}

\begin{table}
  \centering
  \begin{tabular}{|c|c|c|c|c|c|c|}
    \hline
    \multicolumn{3}{|c|}{\multirow{4}{*}{Process}} & \multicolumn{4}{c|}{Number of events - 3L ($\mathcal{L}=4\text{ ab}^{-1}$)}\\
    \cline{4-7}
    \multicolumn{3}{|c|}{\multirow{3}{*}{}} & \vtop{\hbox{\strut No cuts}} & \vtop{\hbox{\strut  $N_b\geq3$}}
    & \vtop{\hbox{\strut  $N_b\geq3$,}\hbox{\strut $H^{\text{lep}}_T>1.6$ TeV}}
    & \vtop{\hbox{\strut  $N_b\geq3$,}\hbox{\strut $H^{\text{lep}}_T>1.6$ TeV}\hbox{\strut $\cancel{E}_T>100$ GeV}}
    
    \\
    \hline
    \multirow{8}{*}{\rotatebox{90}{Signal}} & \multirow{4}{*}{\rotatebox[origin=c]{90}{2-bodies}} & $X\bar{X} \rightarrow \ttWW$ & $137$ & $9$ & $5$ & $4$ \T\\
    & & $T\bar{T} \rightarrow$ \ttHH& $19$ & $5$ & $3$ & $2$ \\
    & & $T\bar{T} \rightarrow$ \ttZH & $35$ & $16$ & $13$ & $11$\\
    & & $T\bar{T} \rightarrow$ \ttZZ& $49$ & $7$ & $6$ & $5$ \B\\
    \cline{2-7}
    & \multirow{4}{*}{\rotatebox[origin=c]{90}{3-bodies}} & $X\bar{X} \rightarrow$ \WWttH& $18$ & $8$ & $5$ & $5$ \T \\
    & & $T\bar{T} \rightarrow$ \WWttH & $15$ & $7$ & $5$ & $4$ \\
    & & $T\bar{T} \rightarrow$ \WWttZ& $5$ & $2$ & $1$ & $1$ \\
    & & $X\bar{X} \rightarrow$ \WWttZ& $4$ & $2$ & $1$ & $1$ \B \\
    \hline
    \multirow{6}{*}{\rotatebox{90}{Background}}&\multicolumn{2}{c|}{\ttW} & $3203$ & $44$ & $2$ & $2$ \T\\
    & \multicolumn{2}{c|}{\ttZ} & $10444$ & $308$ & $13$ & $8$ \\
    & \multicolumn{2}{c|}{\ttH} & $1280$ & $26$ & $0$ & $0$ \\
    & \multicolumn{2}{c|}{\tttt} & $201$ & $79$ & $3$ & $2$ \\    
    & \multicolumn{2}{c|}{\ttWW} & $197$ & $6$ & $0$ & $0$ \\    
    & \multicolumn{2}{c|}{tZ bjj} & $984$ & $21$ & $1$ & $1$ \B \\
    \hline
    \multicolumn{3}{|c|}{$S/B$} & & $0.1$ & $2.1$& $2.6$ \T\B \\
    \hline
    \multicolumn{3}{|c|}{$S/\sqrt{B}$} & & $2.5$ & $9.0$ & $9.3$ \T\B  \\
    \hline
    \end{tabular}
    \caption{Number of events surviving the cuts implementation in the 3L search channel.}
    \label{fig:events_results_3l}
\end{table}

\begin{table}
  \centering
  \begin{tabular}{|c|c|c|c|c|c|c|c|}
    \hline
    \multicolumn{3}{|c|}{\multirow{4}{*}{Process}} & \multicolumn{5}{c|}{Number of events - 2SSL+3L ($\mathcal{L}=4\text{ ab}^{-1}$)}\\
    \cline{4-8}
    \multicolumn{3}{|c|}{\multirow{3}{*}{}} & \vtop{\small \hbox{\strut No} \hbox{\strut cuts}} & \vtop{\small \hbox{\strut  $N_b\geq3$}}
    & \vtop{\footnotesize \hbox{\strut $N_b\geq3$,}\hbox{\strut $H^{\text{lep}}_T>1.6$ TeV}}
    & \vtop{\footnotesize\hbox{\strut  $N_b\geq3$,}\hbox{\strut $H^{\text{lep}}_T>1.6$ TeV}\hbox{\strut $\cancel{E}_T>100$ GeV}}
    & \vtop{\footnotesize \hbox{\strut  $N_b\geq3$,}\hbox{\strut $H^{\text{lep}}_T>1.6$ TeV}\hbox{\strut $\cancel{E}_T>150$ GeV}}
    \\
    \hline
    \multirow{8}{*}{\rotatebox{90}{Signal}} & \multirow{4}{*}{\rotatebox[origin=c]{90}{2-bodies}} & \small $X\bar{X} \rightarrow \ttWW$ & $492$ & $49$ & $32$ & $28$ & $24$ \T\\
    & & \small $T\bar{T} \rightarrow$ \ttHH& $58$ & $19$ & $10$ & $8$ & $7$\\
    & & \small $T\bar{T} \rightarrow$ \ttZH & $52$ & $23$ & $18$ & $14$ & $11$\\
    & & \small $T\bar{T} \rightarrow$ \ttZZ& $60$ & $9$ & $7$ & $6$ & $4$ \B\\
    \cline{2-8}
    & \multirow{4}{*}{\rotatebox[origin=c]{90}{3-bodies}} &  \small $X\bar{X} \rightarrow$ \WWttH& $66$ & $32$ & $24$ & $20$ & $17$ \T \\
    & & \small $T\bar{T} \rightarrow$ \WWttH & $57$ & $27$ & $21$ & $18$ & $15$ \\
    & & \small $T\bar{T} \rightarrow$ \WWttZ& $20$ & $8$ & $6$ & $5$ & $4$ \\
    & & \small $X\bar{X} \rightarrow$ \WWttZ& $16$ & $7$ & $5$ & $4$ & $4$ \B \\
    \hline
    \multirow{6}{*}{\rotatebox{90}{Background}}&\multicolumn{2}{c|}{ \ttW} & $23739$ & $735$ & $35$ & $24$ & $16$ \T\\
    & \multicolumn{2}{c|}{\ttZ} & $17506$ & $545$ & $20$ & $12$ & $8$ \\
    & \multicolumn{2}{c|}{\ttH} & $5174$ & $158$ & $1$ & $1$ & $1$\\
    & \multicolumn{2}{c|}{\tttt} & $859$ & $367$ & $14$ & $10$ & $7$ \\    
    & \multicolumn{2}{c|}{\ttWW} & $794$ & $36$ & $1$ & $1$ & $1$ \\    
    & \multicolumn{2}{c|}{tZ bjj} & $1744$ & $39$ & $1$ & $1$ & $0$ \B \\
    \hline
    \multicolumn{3}{|c|}{$S/B$} & & $0.1$ & $1.7$& $2.1$ & $2.6$ \T\B \\
    \hline
    \multicolumn{3}{|c|}{$S/\sqrt{B}$} & & $4.0$ & $14.5$ & $14.9$ & $15.0$ \T\B  \\
    \hline
    \end{tabular}
    \caption{Number of events surviving the cuts implementation in the 2SSL+3L search channel.}
    \label{fig:events_results_2ssl_3l}
\end{table}

The columns in tables~\ref{fig:events_results_2ssl},~\ref{fig:events_results_3l} and~\ref{fig:events_results_2ssl_3l} show the effect of progressive cuts in the number of events for each of the signal and background channels, as well as the total signal (S) over total background (B) and $S / \sqrt{B}$. Our signal contains 4 b-jets in the final state but most backgrounds do not (the only exception being the four tops background), so the obvious first cut is to demand a number $N_b$ of b-tagged jets, we found that optimal results were obtained for $N_b \ge 3$, which is applied in the three search channels. As the signal is generated from the decay of a pair of heavy particles we expect a lot of transverse momentum to be produced, but in the chosen channels this momentum will be distributed among jets and hard leptons. We thus define $H_T^{lep}$ as the scalar $p_T$ sum of all reconstructed jets and leptons in the event. The minimun $H_T^{lep}$ value turns out to be the most relevant cut in terms of increasing signal to background ratio, and has been optimized to different values for the three channels, as can bee seen in tables~\ref{fig:events_results_2ssl},~\ref{fig:events_results_3l} and~\ref{fig:events_results_2ssl_3l}. Finally we impose a cut on the missing energy ($\cancel{E}_T$) as we also expect neutrinos to be produced from the leptonic decays of the $W$ bosons, and because that will help ensure we are excluding contributions from rarer reducible backgrounds that can generate our signal through miss-identification ($WZ$, $ZZ$, $W^\pm W^\mp$). The optimal values for the  $\cancel{E}_T$ cut are indicated on the tables (we give two possible values for the 2SSL+3L channel).

\section{Conclusions and Outlook}
\label{conclusion}

Vector-like top partners are ubiquitous in models attempting to address the naturalness puzzle of the Standard Model. Using as a concrete example of such an extension, the \five, we have shown that these particles can have sizeable three-body decays, and that taking these channels into account can improve significantly the exclusion limits obtained by previous analyses. Specifically, for the pair production of the lightest top partner from the fourplet state, with one of the legs decaying to $W^+W^-t$, we estimate that the present exclusion limit from CMS in the same-sign dilepton channel \cite{tsearch}, which assumes the width is saturated by two-body decays, would increase from 1.5 TeV up to 1.6 TeV, as shown in figure \ref{fig:T_limit_estimation}. This strongly motivates a more inclusive search for these states to be performed by the experiments.

Although we focused here on the MCHM$_5$, it must be emphasized that we expect these features to be generic in any model containing a vector-like doublet. Most studies so far have 
considered a SM-like doublet (top and bottom partner), with the supposedly conservative assumption of a two-body saturated width. However, we see that, on the contrary, a doublet naturally leads to near degenerate states and sizeable three-body decays. Thus, simplified models built on this assumption are in fact not capturing model independent physics, but instead imposing constraints on their possible UV completions, needed to suppress the three-body channel.

Furthermore, the narrow spectrum leads to large contributions to the production cross section in the three-body decay channels, coming from one of the states being slightly off-shell. The effect can make the cross-section as large as four times the naive estimate from cross-section times branching ratio for narrow states (see figure \ref{fig:distr_x53_fourplet}). This feature makes it difficult to search for one of these states in isolation, hence an inclusive search can be more profitable.

We propose such a search focusing on multileptonic channels (2SSL, 3L and their combination) following the same cut flow used for VLQ searches in these channels. Our results are summarized in tables \ref{fig:events_results_2ssl}, \ref{fig:events_results_3l} and \ref{fig:events_results_2ssl_3l}. With the benchmark point we explored, which predicts a fairly light resonance at 1.3 TeV, the HL-LHC could reach a $S/\sqrt{B}$ of 15.   From this result, and scaling the cross section as $M_{T,X}^{-4}$, we can extrapolate to higher masses and the reach at the HL-LHC would be 1.6 TeV at three sigma, and 1.5 at five sigma. One must notice that this is obtained in a simplified cut-and-count analysis with just two multi-leptonic channels, so a complete analysis will certainly enhance the discovery reach or exclusion potential. Therefore, it is of the utmost importance that the forthcoming analysis implement 3-body decays and an inclusive search for these resonances.

\section{Acknowledgements}
\label{acknowledgements}
The authors thank Geum Bong Yu for useful discussions and references. This work was supported by the S\~ao Paulo Research Foundation
(FAPESP) under grants \#2018/25225-9, \#2021/14335-0. This study was financed in part by the Coordenação de
Aperfeiçoamento de Pessoal de Nível Superior - Brasil (CAPES) - Finance Code 001.
\bibliographystyle{JHEP}
\bibliography{ref}
\end{document}